\documentclass[11pt]{article}

\usepackage[preprint]{acl}

\usepackage{times}
\usepackage{latexsym}
\usepackage{booktabs}
\usepackage{xcolor}
\usepackage{colortbl}
\usepackage{graphicx}
\usepackage{booktabs}
\usepackage{multirow}
\usepackage{xcolor}
\usepackage{pifont}
\usepackage{fontawesome}
\usepackage{amssymb}
\usepackage{enumitem}
\usepackage{booktabs}
\usepackage{multirow}
\usepackage{xcolor}
\usepackage{colortbl}
\usepackage{makecell}
\usepackage{amsmath}
\usepackage[utf8]{inputenc}
\usepackage{booktabs}
\usepackage{tabularx}
\usepackage{multirow}
\usepackage{array}
\usepackage{ragged2e}
\usepackage{geometry}
\usepackage[T1]{fontenc}

\usepackage[utf8]{inputenc}

\usepackage{microtype}

\usepackage{inconsolata}

\usepackage{graphicx}

\usepackage{array}
\newcolumntype{C}[1]{>{\centering\arraybackslash}p{#1}}
\title{MMFCTUB: Multi-Modal Financial Credit Table Understanding Benchmark}

\author{
 \textbf{Cui Yakun\textsuperscript{1}},
 \textbf{Yanting Zhang\textsuperscript{1}},
 \textbf{Zhu Lei\textsuperscript{1}},
 \textbf{Jian Xie\textsuperscript{2}},
\\
 \textbf{Zhizhuo Kou\textsuperscript{1}},
 \textbf{Hang Du\textsuperscript{3}},
 \textbf{Zhenghao Zhu\textsuperscript{1}},
 \textbf{Sirui Han\textsuperscript{1}},
\\
 \textsuperscript{1}The Hong Kong University of Science and Technology,\\
 \textsuperscript{2}DuXiaoman Technology,\\
 \textsuperscript{3}Beijing University of Posts and Telecommunications
\\
 \small{
   \textbf{Correspondence:} \href{Sirui Han}{siruihan@ust.hk}
 }
}

\begin{document}
\maketitle
\begin{abstract}
The advent of multi-modal language models (MLLMs) has spurred research into their application across various table understanding tasks. However, their performance in credit table understanding (CTU) for financial credit review remains largely unexplored due to the following barriers: low data consistency, high annotation costs stemming from domain-specific knowledge and complex calculations, and evaluation paradigm gaps between benchmark and real-world scenarios. To address these challenges, we introduce MMFCTUB (Multi-Modal Financial Credit Table Understanding Benchmark), a practical benchmark, encompassing more than 7,600 high quality CTU samples across 5 table types. MMFCTUB employ a minimally supervised pipeline that adheres to inter-table constraints and maintains data distributions consistency. The benchmark leverages capacity-driven questions and mask-and-recovery strategy to evaluate models' cross-table structure perception, domain knowledge utilization, and numerical calculation capabilities. Utilizing MMFCTUB, we conduct comprehensive evaluations of both proprietary and open-source MLLMs, revealing their strengths and limitations in CTU tasks. MMFCTUB serves as a valuable resource for the research community, facilitating rigorous evaluation of MLLMs in the domain of CTU.

\end{abstract}

\section{Introduction}
Tables have become the predominant format for presenting structured information across various domains due to their superior visual \cite{zheng2024multimodal, zhou2025syntab, kang2025can} accessibility compared to sequential text \cite{wu2025tablebench, wang2024chain, shigarov2023table, deng2022turl}. In financial credit review \cite{brennecke2021commercial}, loan officers need to understand credit report tables(CRTs) and evaluate applicants' economic profiles, which demands accurate cross-table contents recognition, domain knowledge application and numerical computation of assessment metrics.
Multi-modal language models (MLLMs) have recently demonstrated remarkable visual \cite{yang2025qwen3,  wang2025internvl3} understanding capabilities and achieved promising results when applied to image processing tasks across multiple domains \cite{alsaad2024multimodal, ye2024mplug}. Leveraging large-scale pre-training on diverse domain knowledge and mathematical reasoning datasets, MLLMs have developed robust capabilities in domain knowledge integration and numerical computation. These strengths position MLLMs as a promising approach to credit tables understanding (CTU) tasks.
Despite these natural advantages, their capacities have not been comprehensively evaluated in CTU tasks during financial credit review for the following challenges:
\textbf{1. Limited Availability of Data.}Existing table understanding(TU) benchmarks predominantly rely on generic tables collected from public web corpora. However, this approach is inapplicable to CRTs as stringent privacy regulations.
\textbf{2. Specialized Table Dependencies and Distributions}. CRTs exhibit domain-specific layouts with strong cross-table dependencies and consistent distributions reflecting individuals' economic profiles. Existing benchmarks fail to capture these characteristics, thereby introducing data bias into evaluation.
\textbf{3. High Annotation Cost.} Current TU benchmarks predominantly focus on tables with minimal domain knowledge and computational requirements. However, CRTs encode complex economic implications and numerical relations, requiring extensive domain expertise and numerical reasoning during annotation, thus incurring prohibitive costs.
\textbf{4. Misaligned Evaluation Paradigm.} Existing evaluations predominantly process tables in serialized text format. However, this 1D paradigm is inappropriate for practical credit review, where CTU is performed through 2D visual processing, creating a fundamental gap between evaluation and practice.

\begin{table*}[t]
\small
\setlength{\tabcolsep}{3pt}
\begin{tabular}{@{}lcccccccccccccc@{}}
\toprule
\textbf{Dataset} & \textbf{Dom} & \textbf{CrT} & \textbf{CrTPN} & \textbf{CrCDep} & \textbf{CrTDep} & \textbf{TaS} & \textbf{RP} & \textbf{AnC} & \textbf{Par} & \textbf{Kno} & \textbf{Comp} & \textbf{Ann} \\
\midrule
TableBench & Gen & \textcolor{red}{$\times$} & \textcolor{red}{$\times$}  & \textcolor{green}{\checkmark} & \textcolor{red}{$\times$} & Public & \textcolor{red}{$\times$} & High & Text & \textcolor{red}{$\times$} & \textcolor{red}{$\times$} & 19k \\
FewTUD & Gen & \textcolor{red}{$\times$} & \textcolor{red}{$\times$} & \textcolor{green}{\checkmark} & \textcolor{red}{$\times$} & Public &  \textcolor{red}{$\times$} & Mid & Text & \textcolor{red}{$\times$} & \textcolor{red}{$\times$} & 3k\\
Entrant & Fin &  \textcolor{red}{$\times$} &  \textcolor{red}{$\times$}  & \textcolor{green}{\checkmark} &  \textcolor{red}{$\times$} & Public &  \textcolor{red}{$\times$} & Mid & Text &  \textcolor{red}{$\times$} &  \textcolor{red}{$\times$} & 331k\\
FinTab-LLaVA & Fin &  \textcolor{red}{$\times$} & \textcolor{red}{$\times$} & \textcolor{green}{\checkmark} &  \textcolor{red}{$\times$} & Public &  \textcolor{red}{$\times$} & High & Image &\textcolor{green}{\checkmark} & \textcolor{green}{\checkmark} & 7.3k \\
\midrule
\textbf{MMFCTUB} & Fin & \textcolor{green}{\checkmark} & 3 & \textcolor{green}{\checkmark} & \textcolor{green}{\checkmark} & Pub Str+ Syn Data & \textcolor{green}{\checkmark} & Low & Image & \textcolor{green}{\checkmark}  & \textcolor{green}{\checkmark} & 7.7k\\
\bottomrule
\end{tabular}
\captionsetup{justification=raggedright, singlelinecheck=false}
\caption{Comparison of datasets for TU. \textbf{Dom}: Domain. \textbf{CrT}: Cross-Table. \textbf{CrTPN}: Cross Table Paradigm Number, \textbf{CrCDep}: Cross Columns Dependencies, \textbf{CrTDep}: Cross Tables Dependencies, \textbf{TaS}: Table Source. \textbf{RP}: Allign with Credit Review Process. \textbf{AnC}: Annotation Cost. \textbf{Par}: Input Paradigm. \textbf{Kno}: Domain Knowledge. \textbf{Comp}: Computation. \textbf{Anno}: Annotations. \textbf{Gen}: General. \textbf{Fin}: Financial }
\label{tab:dataset_comparison}
\vspace{-12pt}
\end{table*}

To address these challenges, we present MMFCTUB (Multi-Modal Financial Credit Table Understanding Benchmark), a comprehensive benchmark for evaluating MLLMs on table structure perception, domain knowledge utilization and numerical reasoning in CTU tasks. MMFCTUB comprises 19,000 credit table images with authentic layouts across 5 categories, incorporating 60 interdependent fields with coherent per-individual distributions. The benchmark provides 7,600 test instances evaluating 54 financial indicators that characterize applicants' economic profiles, sourced from 246 applicants with diverse economic backgrounds.

During data curation, we employ MLLM-assisted programmatic generation to synthesize credit table images, questions, and annotations with minimal human intervention and near-realistic data. To minimize data bias, we reconstruct CRTs using structure prompts derived from authentic templates and populate cells with values generated from applicants' economic profiles. To preserve inter-field dependencies while ensuring generation efficiency, we adopt a three-tier generation process: LLMs generate descriptive features, while rule-based programs compute high-precision numerical data. CRT images are rendered by compiling LaTeX code generated from abstract table definitions.
During evaluation, MLLMs' are required to generate answers based on table-related questions based on capacity-driven questions, to fine-grained assess MLLMs' knowledge and calculation capacities, we employee a mask-and-recover strategy and hit rate metrics.
Based on MMFCTUB, we conduct a comprehensive evaluation of mainstream MLLMs, including both open-source and proprietary models, as depicted in Table \ref{fintab_data_statics}.
In summary, our contributions are:
\begin{itemize}[itemsep=0pt, parsep=0pt, topsep=0pt]
\item We introduce a MLLM-assisted programmatic generation strategy with low human intervention and high data quality. 
\item We develop MMFCTUB, a novel benchmark for comprehensive CTU, measuring the performance across table structure perception, domain-knowledge utilization and numerical computing.
\item We conduct a comprehensive evaluation of several popular MLLMs, uncovering their strengths and weakness across various dimensions.
\end{itemize}

\section{Related Wrok}
\subsection{Table Understanding}
Recent advances in large language models (LLMs) have spurred significant research in table understanding~\cite{wang2024chain, deng2022turl, chen2024tablerag, cao2025tablemaster}. TableBench~\cite{wu2025tablebench} evaluates LLM performance across 18 domains and four task types using real-world industrial tables. FewTUD~\cite{liu2022few} introduces the first Chinese benchmark for few-shot table understanding. Sui et al.~\cite{sui2024table} adopt a task decomposition approach, assessing LLM capabilities at each intermediate step \cite{li2023making, li2025generator, zhang2025r1, xu2025llava}. However, these works primarily focus on general-purpose tables represented in serialized formats such as Markdown, Pandas DataFrames, or HTML.

\subsection{Financial Table Understanding} 
Comparing with understanding general tables for LLM \cite{lu2025large, chen2024tablerag, ren2025tablegpt}, financial tables are facing with more challenges since they are feature with more complex structures, more domain-specific knowledge \cite{yang2024financial, guo2025fineval, lu2023bbt} and numerical relationships. FinQA \cite{chen2021finqa} extracts tables from financial market reports and provides a dataset for numerical calculation \cite{su2024numllm, loukas2022finer, khang2024data} tasks. \cite{zhu2021tat} TAT-QA addresses the unique requirements of financial tables by defining subtasks including sequence tagging, aggregation operators, and scale prediction to evaluate LLM performance. 
\subsection{MLLMs for Table Understanding Benchmark} 
Multimodal large language models (MLLMs) \cite{yang2025qwen3, wang2025internvl3} have been widely applied across domains \cite{tampubolon2025challenges, chen2025multimodal}, prompting domain-specific benchmarks to evaluate their capabilities. Financial tasks require domain knowledge, numerical reasoning, and table structure perception. \cite{zheng2024multimodal} proposes Table-LLaVA, processing table images directly with strong performance across 23 benchmarks. FinTab-LLaVA \cite{park2025fintab} introduces a multimodal LLM tuned on FinTMD for financial table QA, fact verification, and description generation. Credit reports encode economic profiles through specialized layouts and numerical relationships critical for loan decisions, yet existing benchmarks inadequately assess MLLM performance on such tables. We present MMFCTUB, a comprehensive benchmark evaluating MLLMs on financial credit report understanding.
\section{MMFCTUB}
In this section, we deleve into the meticulous construction of MMFCTUB, introducing our strategic approach to innovative methods employed to generation simulation credit report tables data, design of a comprehensive capability taxonomy, innovative methods leveraged to assess specific capacities, evaluation metrics definition. Additionally, we present detailed statistics of MMFCTUB
and contrast it with existing finance table benchmarks, tthereby illustrating its unique features and contributions to the field.
\begin{figure*}[!htbp]  
    \centering
    \includegraphics[width=\textwidth]{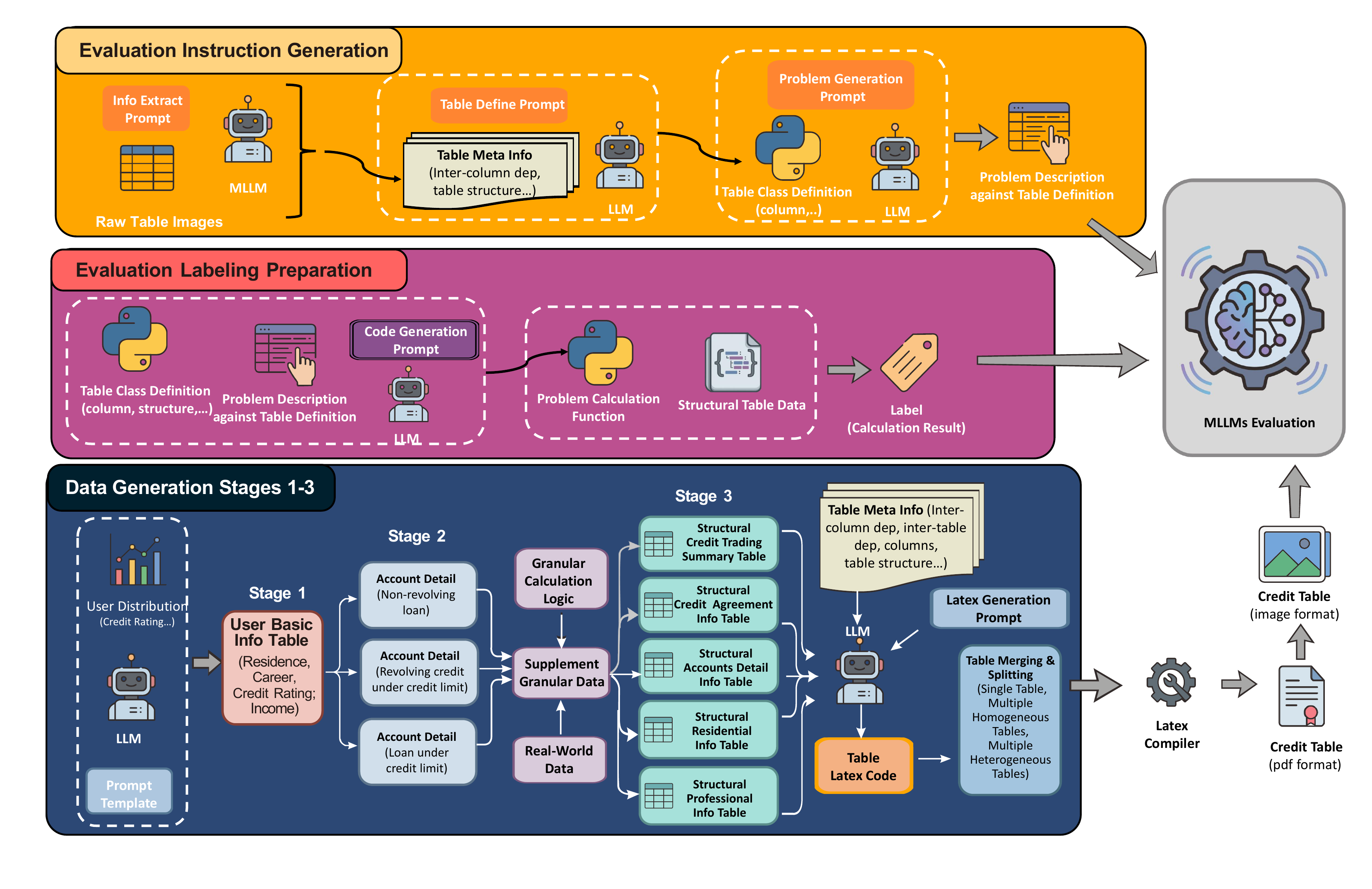}  
    \captionsetup{justification=centering, singlelinecheck=false}
    \caption{Detail of Finance Credit Table Understanding Dataset Construction.}  
    \label{data_construct_fintab} 
\end{figure*}

\subsection{Benchmark Construction}
\label{label_bench_contsuction}
\begin{figure}[t]  
    \hspace{-10pt}
    \includegraphics[scale=0.35]{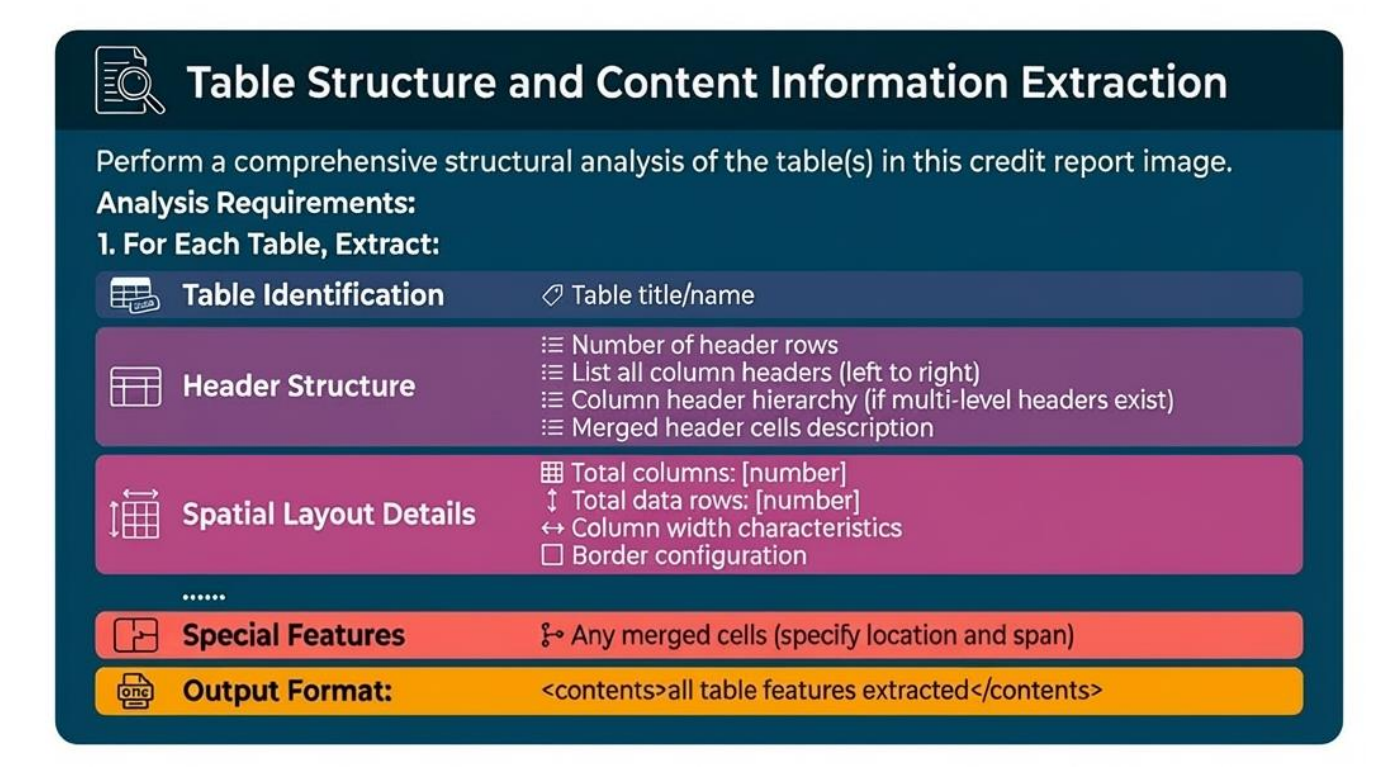}  
    \captionsetup{justification=centering, singlelinecheck=false}
    \caption{Extracted Table Contents and Meta Info.}
    \label{data_construct_table_contentx_extract}
    \vspace{-12pt}
\end{figure}

\begin{table}[t]
\centering
\scriptsize 
\label{tab:cross_table_paradigm_single_column} 
\begin{tabular}{@{}l l@{}}
\toprule
\textbf{Cross-table Paradigm} & \textbf{Corresponding Table Types} \\
\midrule
Cross-table Between Homo Static Table & Credit Agreement Table \\
\addlinespace[0.3em] 
Cross-table Between Homo Dynamic Table & \makecell[l]{Credit Trans. Table, \\ Acct. Detail Table} \\
\addlinespace[0.3em]
Cross-table Between Hetero Table & \makecell[l]{Residence + Acct., \\ Occupation + Acct.} \\
\bottomrule
\end{tabular}
\caption{Correspondence Between Three Cross-table Paradigms and Table Types. Homo. = Homogeneous. Hetero = Heterogeneous. Trans. = Transaction; Acct. = Account; Credit Trans. = Credit Transaction; Residence + Acct. = Residence Information and Account Detail; Occupation + Acct. = Occupation Information and Account Detail.}
\label{table_paradigm_and_type}
\vspace{-12pt}
\end{table}

\begin{figure}[t]  
    \hspace{-25pt}
    \includegraphics[scale=0.42]{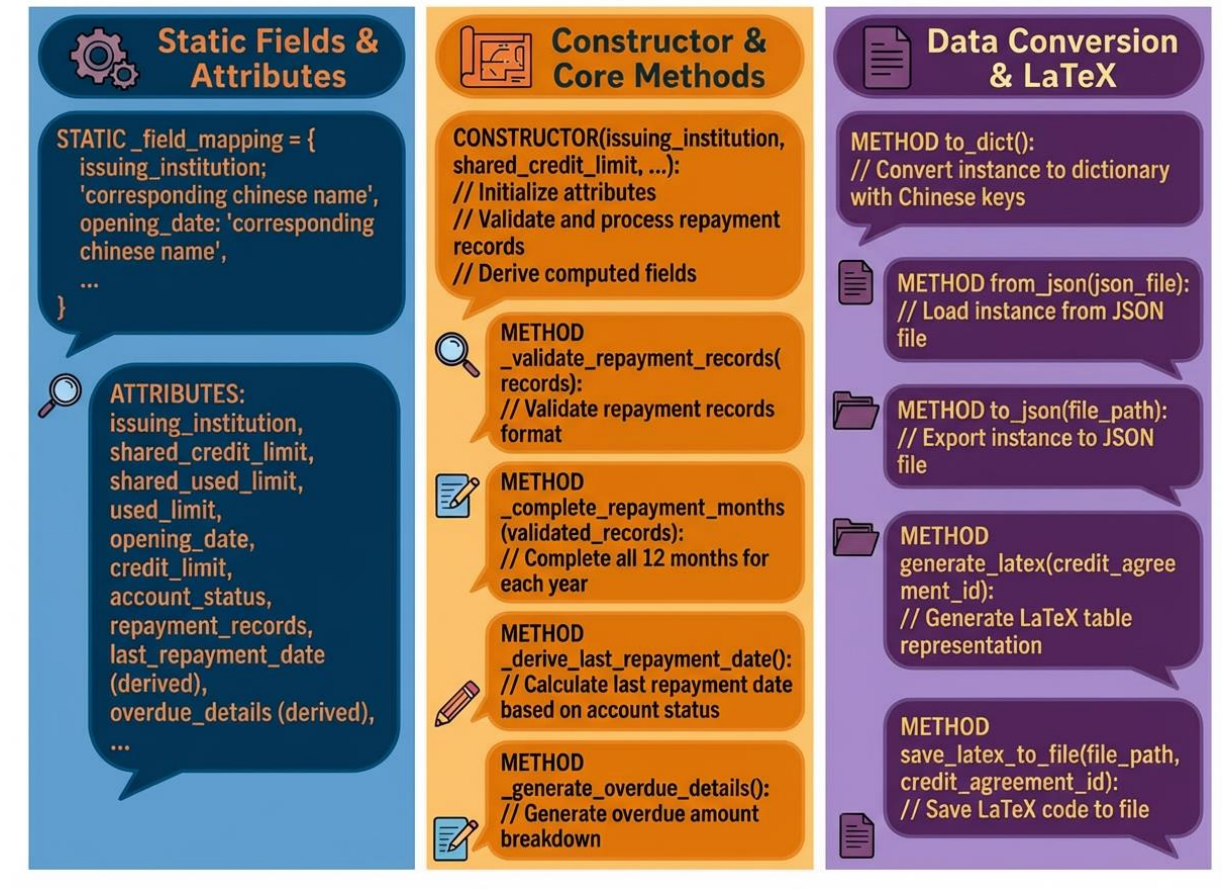}  
    \captionsetup{justification=raggedright, singlelinecheck=false}
    \caption{Detail of Definition of Credit Card.}
    \label{data_construction_function_demo}
    \vspace{-12pt}
\end{figure}

\textbf{Credit Table Selection.} In practical loan review processes, reviewers must scan multiple relevant tables and extract target data for computation, requiring frequent cross-table understanding. We therefore adopt a cross-table input paradigm for our benchmark. Based on consultation with domain experts, we select 5 commonly used table types: credit transaction tables, residence information tables, occupation information tables, account detail tables, and credit agreement tables.\\
\textbf{Credit Table Generation}. 
Existing table generation approaches suffer from two critical limitations: (1)Failing to capture the specific characteristics and requirements of domain-specific scenarios, necessitating substantial efforts in data collection and cleaning, and (2) prohibitively high annotation costs when domain experts are involved in large-scale and complex labeling. To address these dataset limitations, We propose a novel credit table generation methodology that aligns with practical credit review processes while requiring minimal human effort. As illustrated in Figure \ref{data_construct_fintab}, our approach comprises three key components: \\
\hspace*{1em}\textbf{1) Instruction Generation.} Credit tables exhibit domain-specific layouts and dependencies defined by experts, represented through table metadata. Our data construction pipeline involves 3 steps. (1) extracting metadata from real-world credit tables using MLLMs, (2) constructing abstract table representations via LLM based on semantic metadata (Figure \ref{data_construction_function_demo}), and (3) generating questions targeting credit review indicators through LLM with expert-designed prompts.\\
\hspace*{1em}\textbf{2) Labels Preparation.} Credit table computations involve domain-specific economic indicators that require complex multi-column and cross-table aggregations, rendering manual annotation impractical. To mitigate this, we leverage LLMs to generate label calculation function code from abstract table definitions and questions. Ground truth labels are obtained by executing these functions on the generated table data.\\
\hspace*{1em}\textbf{3) Data Construction.} Credit tables exhibit complex inter-column and cross-table dependencies aligned with applicants' economic profiles. We employ a three-stage generation pipeline. Stage 1: LLMs generate user basic information from diverse economic profiles defined by income and credit score distributions. Stage 2: LLMs produce account detail tables using dependency-aware prompts, followed by rule-based population of numerically constrained fields using real-world financial parameters. Stage 3: Summary and agreement tables are generated from account tables to maintain cross-table dependencies. Tables are rendered as images via LLM-generated LaTeX code for visual QA tasks. This pipeline ensures structural authenticity, semantic coherence, and preservation of complex interdependencies characteristic of real credit data.\\
\textbf{Capacity Taxonomy.} Drawing inspiration from the operational workflows of credit reviewers in real-world assessment processes, we propose three task categories to evaluate MLLMs' credit table understanding capabilities: Table Structure Perception (TSP), Domain Knowledge Utilization (DKU), and Numerical Calculation (NC).

\textbf{TSP} assesses models' capacity to perceive table structures, including spatial layouts and cross-table relationships. Evaluation spans two dimensions: structural perception across multiple paradigms (Table \ref{table_paradigm_and_type}), and scope perception across three table count ranges (3-5, 6-8, 9-13). By minimizing domain knowledge and computational demands in question design, we isolate structural perception capabilities. Performance is measured by answer accuracy.

\textbf{DKU}, assesses MLLMs' capacity to apply pre-trained financial knowledge, including terminology, relationships, and formulas. Models must construct formulations and select variables based on entity relationships within perceived table structures. We categorize domain knowledge difficulty into three levels:
We define three knowledge difficulty levels. \textbf{Knowledge Perception} requires extracting information directly present in tables, such as field names in headers. \textbf{Knowledge Analysis} demands deriving information through joint analysis of question semantics and table content, such as computing field-level sums or ratios. \textbf{Knowledge Reasoning} requires applying knowledge from prompt context independent of table content, where models must comprehend logical flows and infer intermediate variables for subsequent calculations.

\textbf{NC}, MLLMs' calculation capacity in CTU tasks derives from general arithmetic reasoning and domain-specific numerical relationship understanding. We evaluate three operators frequently used in CTU scenarios: addition (+) for aggregating values across tables (e.g., summing column values from different account tables), subtraction (-) for computing duration metrics (e.g., days between account opening and closure), and division (÷) for calculating proportional metrics (e.g., credit utilization ratios measuring account borrowing saturation).
\\
\textbf{Evaluation Paradigm.}We establish distinct evaluation paradigms for three CTU capacities, using differentiated questions to isolate specific capability bottlenecks. For \textbf{structure perception}, questions require extensive table extraction with minimal computation, measuring accuracy on unmasked samples to isolate structural understanding independent of other capabilities. \\
For \textbf{domain knowledge utilization}, we employ mask-and-recovery evaluation where column names are randomly masked with 'XXXX' tokens. We use Financial Knowledge Hit Rate (FKHR) to measure proficiency in identifying and applying domain-specific knowledge.
\begin{equation}
\text{FKHR}_{ij} = \frac{|P_{ij} \cap K_{ij}|}{|K_{ij}|},
\end{equation}
where $P_{ij}$ and $K_{ij}$ denote the predicted and label knowledge groups for question $i$ and group $j$. The term $|P_{ij} \cap K_{ij}|$ counts correctly identified elements, while $|K_{ij}|$ is the ground truth size. The metric equals 1 when all elements are correctly predicted.
\begin{equation}
\overline{\text{FKHR}}_j = \frac{1}{N} \sum_{i=1}^{N} \text{FKHR}_{ij},
\end{equation}
where $N$ denotes the total number of questions in the dataset, and $\text{FKHR}_{ij}$ represents the hit rate for the $i$-th question on the $j$-th knowledge group. This metric reflects the model's overall prediction performance on a specific knowledge group.
\begin{figure*}[!htbp]  
    \centering
    \includegraphics[width=\textwidth]{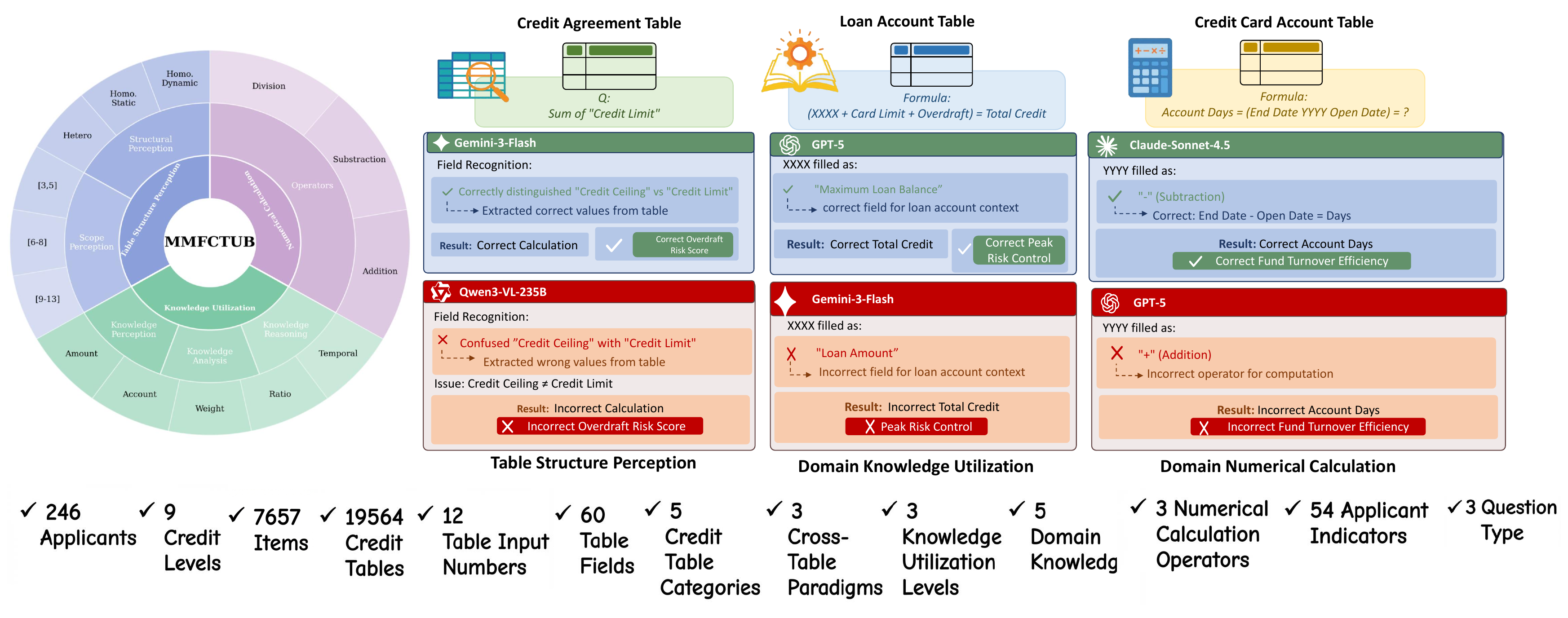}  
    \captionsetup{justification=centering, singlelinecheck=false}
    \caption{The Comprehensive Taxonomy, Data Examples and Statistical Characteristics of MMFCTUB. The circular
taxonomy diagram shows core cognitive levels, knowledge categories and operators.}  
    \label{data_all_show} 
\end{figure*}
\begin{figure*}[!htbp]  
    \centering
    \includegraphics[width=\textwidth]{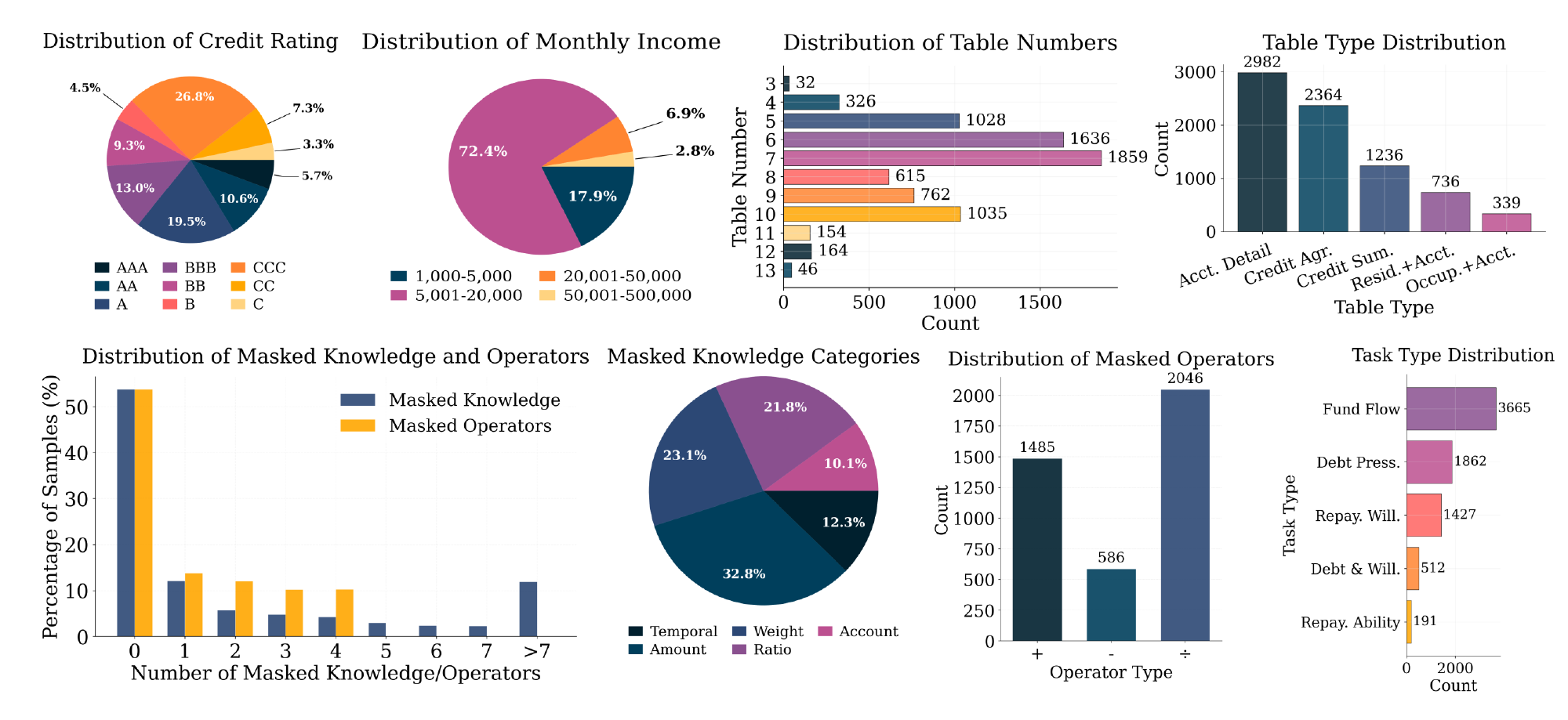}  
    \captionsetup{justification=centering, singlelinecheck=false}
    \caption{Dataset Details of MMFCTUB.}  
    \label{fintab_data_statics} 
\end{figure*}
For numerical calculation, we assess MLLMs through the mask-and-recovery paradigm. During question generation, the program randomly masks calculation operators from an expert-predefined list in the prompts using 'YYYY' as the mask token. During evaluation, we employ calculation operator hit rate (COHR) as the metric, quantifying MLLMs' proficiency in identifying and applying appropriate operators essential to the calculation process:
\begin{equation}
\text{COHR}_{ij} = \frac{1}{|O_{ij}|} \sum_{k=1}^{|O_{ij}|} \mathbb{1}(O_{ijk} = P_{ijk}),
\label{eq:cohr}
\end{equation}

\noindent where $O_{ij} = \{O_{ij1}, O_{ij2}, \ldots, O_{ijn}\}$ denotes the ground truth operator sequence for the $j$-th operator group in the $i$-th question, and $P_{ij} = \{P_{ij1}, P_{ij2}, \ldots, P_{ijm}\}$ represents the corresponding predicted sequence. The indicator function $\mathbb{1}(\cdot)$ returns 1 if the condition is satisfied and 0 otherwise. $|O_{ij}|$ denotes the length of the ground truth sequence.

Unlike the knowledge hit rate which uses set intersection, COHR enforces strict positional matching: an operator at position $k$ is considered correct only if $O_{ijk} = P_{ijk}$. When the predicted sequence is shorter than the ground truth ($k > |P_{ij}|$), the missing positions are counted as mismatches.
\begin{equation}
\overline{\text{COHR}}_j = \frac{1}{N} \sum_{i=1}^{N} \text{COHR}_{ij},
\label{eq:cohr_avg}
\end{equation}

\noindent where $N$ denotes the total number of questions in the dataset. This metric reflects the model's overall performance on a specific operator group.
\subsection{Dataset Statistics}
MMFCTUB contains 246 users characterized by credit ratings and monthly income, with equal distributions across three rating ranges and realistic income distributions (Figure \ref{fintab_data_statics}). The dataset comprises 19k tables across multiple types, where table numbers and categories are determined by user economic profiles. All distributions are shown in Figure \ref{fintab_data_statics}.

The benchmark comprises 18k training and 7.6k test QA pairs across three formats (calculation, single-choice, multiple-choice), each targeting specific MLLM capabilities. To evaluate knowledge utilization and numerical reasoning, we randomly mask predefined knowledge elements and computational operators. Evaluation covers five credit assessment knowledge categories (temporal, amount, account, weight, ratio), each requiring distinct table content scopes and domain reasoning. Knowledge and operator distributions are presented in Figure \ref{fintab_data_statics}.
\section{Experiment}
Utilizing MMFCTUB, we conduct a comprehensive evaluation of diverse MLLMs, encompassing both open-source and proprietary systems. During evaluation, we employ the default hyperparameters specified in their respective official implementations for inference. We leverage LaTeX code generation coupled with compilation tools to convert structured tables into image format, with each table rendered as an independent image file for visual processing by the models.
\subsection{Main Results}
\begin{table*}[!t]
\centering
\renewcommand{\arraystretch}{1.1}
\setlength{\tabcolsep}{3pt}
\resizebox{\linewidth}{!}{%
\begin{tabular}{lccccccccccccc}
\toprule[1.2pt]
\rowcolor{gray!20}
\textbf{MLLMs} & \multicolumn{3}{c}{\cellcolor{blue!15}\textbf{Structure Perception}} & \multicolumn{3}{c}{\cellcolor{green!15}\textbf{Scope Perception}} & \multicolumn{3}{c}{\cellcolor{orange!15}\textbf{Knowledge Utilization}} & \multicolumn{3}{c}{\cellcolor{purple!15}\textbf{Numerical Calculation}} & \cellcolor{pink!15}\textbf{Final} \\
\cmidrule(lr){2-4} \cmidrule(lr){5-7} \cmidrule(lr){8-10} \cmidrule(lr){11-13}
& \cellcolor{blue!15}\textbf{\makecell{Homo.\\Static}} & \cellcolor{blue!15}\textbf{\makecell{Homo.\\Dynamic}} & \cellcolor{blue!15}\textbf{\makecell{Hetero.}} & \cellcolor{green!15}\textbf{[3, 5]} & \cellcolor{green!15}\textbf{[6, 8]} & \cellcolor{green!15}\textbf{[9, 14]} & \cellcolor{orange!15}\textbf{\makecell{Know.\\Percep.}} & \cellcolor{orange!15}\textbf{\makecell{Know.\\Analy.}} & \cellcolor{orange!15}\textbf{\makecell{Know.\\Reason.}} & \cellcolor{purple!15}\textbf{Add.} & \cellcolor{purple!15}\textbf{Sub.} & \cellcolor{purple!15}\textbf{Div.} & \cellcolor{pink!15}\textbf{Acc.} \\
\midrule[1pt]
\rowcolor{yellow!20}
\multicolumn{14}{l}{\textit{Proprietary Models}} \\
\midrule[0.5pt]
Gemini-3-Flash & \cellcolor{red!25}81.70 & \cellcolor{red!25}76.15 & \cellcolor{red!25}63.92 & \cellcolor{red!25}76.81 & \cellcolor{red!25}76.63 & \cellcolor{red!25}80.69 & \cellcolor{red!15}41.47 & \cellcolor{red!15}34.81 & 24.87 & \cellcolor{red!15}43.83 & \cellcolor{red!25}24.12 & \cellcolor{red!15}52.68 & \cellcolor{red!25}78.18 \\
Sonnet4.5 & 59.76 & 50.84 & 33.39 & 65.19 & 59.68 & 53.01 & 31.96 & 31.31 & 30.33 & 44.91 & 18.67 & 50.88 & 52.96 \\
Sonnet4.5-think & 69.68 & 56.70 & \cellcolor{red!15}41.26 & 66.34 & 59.89 & 58.67 & 37.27 & 34.26 & \cellcolor{red!25}31.45 & \cellcolor{red!25}44.88 & \cellcolor{red!15}18.02 & \cellcolor{red!25}54.95 & 56.86 \\
GPT-5 & \cellcolor{red!15}77.83 & \cellcolor{red!15}58.41 & 31.50 & \cellcolor{red!15}68.38 & \cellcolor{red!15}63.82 & \cellcolor{red!15}64.33 & \cellcolor{red!25}52.84 & \cellcolor{red!25}36.71 & \cellcolor{red!15}27.98 & 22.47 & 2.92 & 22.06 & \cellcolor{red!15}63.68 \\
GPT-4o & 27.77 & 27.82 & 31.59 & 31.16 & 28.07 & 25.44 & 14.32 & 9.53 & 9.49 & 16.31 & 16.31 & 13.48 & 31.01 \\
\midrule[1pt]
\rowcolor{cyan!20}
\multicolumn{14}{l}{\textit{Open-source Models}} \\
\midrule[0.5pt]
GLM-4\_6V & 38.81 & 52.99 & 37.10 & 45.30 & 43.29 & 40.94 & 8.23 & 8.39 & 5.28 & 9.31 & 4.21 & \cellcolor{red!15}13.37 & 43.93 \\
GLM-4\_1V  & 45.04 & 50.65 & 35.86 & \cellcolor{red!15}57.48 & \cellcolor{red!15}50.59 & 43.81 & 12.07 & 7.39 & 5.78 & 6.08 & 3.81 & 9.87 & 45.75 \\
Keye-VL-1.5-8B & 44.20 & 39.66 & 25.18 & 52.44 & 49.28 & 39.62 & 4.75 & 4.19 & 4.13 & 6.97 & 2.12 & 10.39 & 40.11 \\
Keye-VL-8B  & 45.24 & 40.98 & 33.85 & 49.53 & 50.11 & 41.30 & 7.74 & 8.31 & 8.34 & 12.11 & 2.5 & 10.77 & 42.09 \\
MiniCPM-V-4\_5 & 39.65 & 30.50 & 26.10 & 44.28 & 38.64 & 29.49 & 11.64 & 2.97 & 4.62 & 4.56 & 1.46 & 5.37 & 34.01 \\
InternVL3\_5-8B & 49.37 & 46.45 & 36.59 & 51.66 & 48.59 & 43.39 & 10.04 & 5.02 & 3.66 & 4.66 & 2.54 & 5.98 & 46.37 \\
InternVL3\_5-38B & 49.83 & 45.74 & 35.12 & 51.83 & 47.55 & 44.68 & 10.83 & 5.13 & 4.10 & 4.71 & 3.25 & 6.18 & 47.39 \\
InternVL3\_5-241B-A28B & \cellcolor{red!15}50.14 & \cellcolor{red!15}55.21 & \cellcolor{red!25}38.57 & 52.55 & 48.61 & \cellcolor{red!15}45.69 & 12.56 & 5.21 & 4.85 & 4.61 &3.82 & 6.55 & \cellcolor{red!15}47.55 \\
Qwen3-VL-8B & 37.39 & 53.18 & 32.30 & 36.82 & 38.50 & 39.40 & 15.31 & 5.41 & 6.64 & 14.19 & 4.94 & 5.88 & 42.63 \\
Qwen3-VL-8B-think & 46.73 & 40.30 & 31.08 & 42.91 & 44.55 & 36.22 & 10.30 & 10.8 & 7.70 & \cellcolor{red!15}14.98 & 4.57 & 6.36 & 43.57 \\
Qwen3-VL-30B-think & 54.46 & 42.55 & 36.37 & 44.15 & 45.09 & 37.11 & 11.34 & \cellcolor{red!15}11.05 & \cellcolor{red!15}7.97 & 14.56 & \cellcolor{red!15}5.63 & 7.15 & 45.92 \\
Qwen3-VL-235B-think & \cellcolor{red!25}59.52 & \cellcolor{red!25}65.00 & \cellcolor{red!15}37.11 & \cellcolor{red!25}60.67 & \cellcolor{red!25}61.32 & \cellcolor{red!25}58.40 & \cellcolor{red!25}25.47 & \cellcolor{red!25}24.48 & \cellcolor{red!25}22.89 & \cellcolor{red!25}24.95 & \cellcolor{red!25}15.10 & \cellcolor{red!25}29.15 & \cellcolor{red!25}55.49 \\
Qwen2.5-VL-72B-Instruct & 28.38 & 24.26 & 26.14 & 24.48 & 23.41 & 29.69 & \cellcolor{red!15}21.53 & 5.23 & 9.99 & 10.23 & 5.16 & 6.25 & 26.69 \\
\bottomrule[1.2pt]
\end{tabular}%
}
\caption{\textbf{Performance Comparison of MLLMs Across Different Evaluation Dimensions.} Under Structure Perception: Homo. Static = Cross-table Operations Between Homogeneous Static Tables; Homo. Dynamic = Cross-table Operations Between Homogeneous Dynamic Tables; Hetero.= Cross-table Operations Between Heterogeneous Tables. Under Knowledge Utilization: Know. Percep. = Knowledge Perception; Know. Analy. = Knowledge Analysis; Know. Reason. = Knowledge Reasoning. Under Numerical Calculation: Add. = Addition; Sub. = Subtraction; Div. = Division. \cellcolor{red!25}Deep red indicates the highest value in each column, \cellcolor{red!15}light red indicates the second highest.}
\label{tab:mllm_performance}
\end{table*}

\begin{table*}[!t]
\centering
\renewcommand{\arraystretch}{1.1}
\setlength{\tabcolsep}{2.5pt}
\resizebox{\linewidth}{!}{%
\begin{tabular}{lcccccccccccccccc}
\toprule[1.2pt]
\rowcolor{gray!20}
\textbf{MLLMs} & \multicolumn{5}{c}{\cellcolor{blue!15}\textbf{Table Type}} & \multicolumn{5}{c}{\cellcolor{green!15}\textbf{Domain Knowledge}} & \multicolumn{5}{c}{\cellcolor{orange!15}\textbf{Task Type}} & \cellcolor{pink!15}\textbf{Final Acc} \\
\cmidrule(lr){2-6} \cmidrule(lr){7-11} \cmidrule(lr){12-16}
& \cellcolor{blue!15}\textbf{\makecell{Credit\\Trans.}} & \cellcolor{blue!15}\textbf{\makecell{Residence\\+ Acct.}} & \cellcolor{blue!15}\textbf{\makecell{Credit\\Agreement}} & \cellcolor{blue!15}\textbf{\makecell{Occupation\\+ Acct.}} & \cellcolor{blue!15}\textbf{\makecell{Account\\Details}} & \cellcolor{green!15}\textbf{Temporal} & \cellcolor{green!15}\textbf{Amount} & \cellcolor{green!15}\textbf{Account} & \cellcolor{green!15}\textbf{Weight} & \cellcolor{green!15}\textbf{Ratio} & \cellcolor{orange!15}\textbf{\makecell{Fund\\Flow}} & \cellcolor{orange!15}\textbf{\makecell{Debt\\Pressure}} & \cellcolor{orange!15}\textbf{\makecell{Repay.\\Willing.}} & \cellcolor{orange!15}\textbf{\makecell{Debt \&\\Willing.}} & \cellcolor{orange!15}\textbf{\makecell{Repay.\\Ability}} & \\
\midrule[1pt]
\rowcolor{yellow!20}
\multicolumn{17}{l}{\textit{Proprietary Models}} \\
\midrule[0.5pt]
Gemini-3-Flash & \cellcolor{red!15}58.90 & \cellcolor{red!25}56.06 & \cellcolor{red!25}81.70 & \cellcolor{red!25}80.72 & \cellcolor{red!25}89.47 & 13.76 & \cellcolor{red!25}45.03 & \cellcolor{red!25}37.50 & \cellcolor{red!15}30.64 & 15.96 & \cellcolor{red!25}77.58 & \cellcolor{red!25}74.87 & \cellcolor{red!25}87.73 & \cellcolor{red!25}63.64 & \cellcolor{red!25}88.54 & \cellcolor{red!25}78.18 \\
Sonnet4.5 & 47.52 & 35.00 & 65.89 & 29.94 & 50.84 & 11.80 & 35.95 & 30.74 & 22.46 & 36.33 & 57.26 & 48.23 & 49.50 & 55.14 & 32.35 & 52.96 \\
Sonnet4.5-think & \cellcolor{red!25}51.34 & \cellcolor{red!15}40.28 & 69.68 & 43.40 & 62.62 & \cellcolor{red!25}37.50 & \cellcolor{red!25}41.97 & 29.07 & \cellcolor{red!25}37.37 & \cellcolor{red!15}37.37 & 59.53 & 52.22 & 38.65 & \cellcolor{red!15}55.41 & 38.65 & 56.86 \\
GPT-5 & 39.63 & 24.83 & \cellcolor{red!15}77.83 & \cellcolor{red!15}45.89 & \cellcolor{red!15}73.10 & \cellcolor{red!15}20.47 & 20.47 & \cellcolor{red!15}32.05 & 18.88 & \cellcolor{red!25}38.64 & \cellcolor{red!15}61.58 & \cellcolor{red!15}61.58 & \cellcolor{red!15}66.49 & 44.19 & \cellcolor{red!15}76.47 & \cellcolor{red!15}63.68 \\
GPT-4o & 20.15 & 28.93 & 27.77 & 37.34 & 33.78 & 3.79 & 13.16 & 7.12 & 10.33 & 6.18 & 24.78 & 34.96 & 43.55 & 26.45 & 31.62 & 31.01 \\
\midrule[1pt]
\rowcolor{cyan!20}
\multicolumn{17}{l}{\textit{Open-source Models}} \\
\midrule[0.5pt]
GLM-4\_6V & 34.58 & \cellcolor{red!15}37.70 & 38.81 & \cellcolor{red!15}35.96 & \cellcolor{red!15}52.99 & 2.05 & 1.05 & \cellcolor{red!25}9.79 & 7.44 & 3.50 & 41.97 & 45.84 & 47.97 & 39.81 & 43.88 & 43.93 \\
GLM-4\_1V & 31.31 & \cellcolor{red!25}38.55 & 45.04 & 30.70 & 50.65 & 2.95 & 11.33 & 9.13 & 7.31 & 3.17 & \cellcolor{red!15}45.13 & 49.70 & 47.33 & 34
.95& 35.20 & 45.75 \\
Keye-VL-1.5-8B & 30.43 & 23.83 & 44.20 & 27.70 & 39.66 & 1.11 & 5.08 & 5.41 & 5.05 & 2.01 & 38.67 & 45.44 & 39.78 & 35.39 & 32.26 & 40.11 \\
Keye-VL-8B & 37.71 & 32.55 & 45.24 & 36.41 & 40.98 & 3.43 & 9.56 & 7.46 & \cellcolor{red!25}9.20 & 4.82 & 41.33 & 44.88 & 40.74 & 45.72 & 27.91 & 42.09 \\
MiniCPM-V-4\_5 & 27.18 & 27.83 & 39.65 & 27.83 & 30.05 & 2.42 & 6.02 & 4.69 & 5.82 & 2.27 & 33.35 & 36.95 & 33.94 & 30.17 & 27.46 & 34.01 \\
InternVL3\_5-8B & 34.81 & 35.99 & 49.37 & 37.75 & 46.45 & 2.64 & 8.13 & 5.39 & 5.14 & 1.97 & 43.21 & 50.46 & 50.24 & 43.55 & 45.92 & 46.37 \\
InternVL3\_5-38B & 35.16 & 36.74 & 49.83 & 37.16 & 46.37 & 2.75 & 8.55 & 5.53 & 4.06 & 2.16 & 44.51 & \cellcolor{red!15}50.95 & 51.32 & 44.15 & 46.31 & 47.39 \\
InternVL3\_5-241B-A28B & \cellcolor{red!25}36.25 & 35.74 & \cellcolor{red!15}50.14 & 38.15 & 47.12 & \cellcolor{red!15}4.85 & \cellcolor{red!15}14.71 & 6.74 & \cellcolor{red!15}9.12 & \cellcolor{red!15}9.27 & 43.56 & 48.53 & \cellcolor{red!15}52.34 & \cellcolor{red!15}45.11 & \cellcolor{red!15}47.10 & \cellcolor{red!15}47.55 \\
Qwen3-VL-8B & 30.97 & 31.31 & 37.39 & 34.19 & 53.18 & 3.19 & 11.64 & 1.92 & 8.26 & 2.23 & 38.26 & 49.17 & 46.44 & 39.77 & 41.80 & 42.63 \\
Qwen3-VL-8B-think & 30.97 & 31.31 & 46.73 & 34.19 & \cellcolor{red!25}53.18 & 3.23 & 11.89 & 2.13 & 8.51 & 2.53 & 39.62 & 44.81 & 44.60 & 35.70 & \cellcolor{red!25}52.55 & 43.57 \\
Qwen3-VL-30B-think & 31.70 & 29.82 & 54.46 & 33.49 & 40.30 & 1.93 & 13.08 & 3.29 & 6.98 & 2.63 & 39.54 & 43.98 & 40.15 & 29.49 & 32.98 & 45.92 \\
Qwen3-VL-235B-think & \cellcolor{red!15}36.08 & 30.36 & \cellcolor{red!25}59.52 & \cellcolor{red!25}39.71 & 42.72 & \cellcolor{red!25}5.14 & \cellcolor{red!25}15.74 & \cellcolor{red!15}6.82 & 8.82 & \cellcolor{red!25}9.48 & \cellcolor{red!25}53.23 & \cellcolor{red!25}60.89 & \cellcolor{red!25}58.29 & \cellcolor{red!25}52.44 & 32.89 & \cellcolor{red!25}55.40 \\
Qwen2.5-VL-72B-Instruct & 14.16 & 22.42 & 28.38 & 34.22 & 31.72 & 3.15 & 10.37 & 5.17 & 7.26 & 4.68 & 21.69 & 34.75 & 32.45 & 16.21 & 29.32 & 26.69 \\
\bottomrule[1.2pt]
\end{tabular}%
}
\caption{\textbf{Performance Comparison of MLLMs Across Different Table Types, Domain Knowledge, and Task Types.} Under Table Type: Credit Trans. = Credit Transaction; Residence + Acct. = Residence and Account; Occupation + Acct. = Occupation and Account. Under Task Type: Repay. Willing. = Repayment Willingness; Debt \& Willing. = Debt and Willingness; Repay. Ability = Repayment Ability. \cellcolor{red!25}Deep red indicates the highest value in each column, \cellcolor{red!15}light red indicates the second highest.}
\label{tab:mllm_performance_extended}
\end{table*}

\begin{figure}[t]
  \includegraphics[width=\columnwidth]{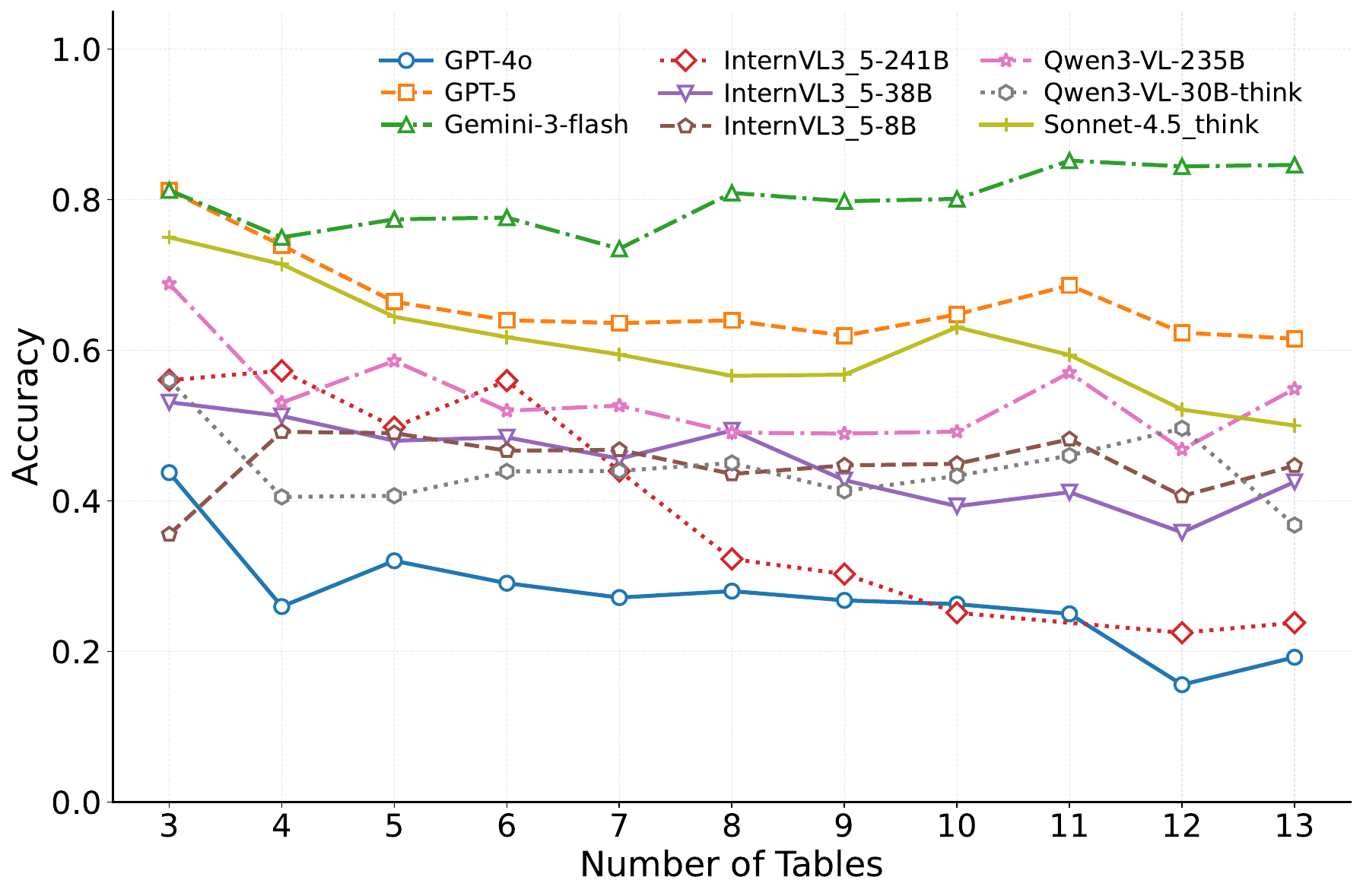}
  \caption{Comparison of Credit Tables Understanding with Number of Tables.}
  \label{model_accuracy_vs_table_num}
  \vspace{-12pt}
\end{figure}

\textbf{Open-Source MLLMs.} We evaluate several prominent open-source MLLMs that support multi-image inference: Qwen-VL-2.5 \cite{bai2025qwen2}, Qwen-VL-3 \cite{yang2025qwen3}, Intern-VL-3 \cite{wang2025internvl3}, and Intern-VL-3.5 \cite{wang2025internvl3}. Notably, Qwen3-VL-235B-think achieves competitive performance, surpassing GPT-4o \cite{islam2025gpt} by 24\% in accuracy, particularly excelling in domain knowledge utilization and numerical calculation. Interestingly, within the same model family, smaller models demonstrate superior performance on complex table understanding tasks.\\
\textbf{Proprietary MLLMs.} We evaluate four proprietary MLLMs: Gemini-3-Flash \cite{comanici2025gemini}, Claude Sonnet 4.5 \cite{salbas2025performance}, GPT-5 \cite{wang2025capabilities}, and GPT-4o. Gemini-3-Flash leads with 15\% higher accuracy than the second-best model, excelling in structure recognition and scope perception. GPT-5 shows superior knowledge utilization across all capability levels, while Sonnet 4.5 achieves best numerical computation performance. GPT-4o underperforms at 31\% accuracy. We attribute these results to the distinct capability profiles of each model.  Notably, GPT-5 achieves second-best overall performance despite weaker numerical calculation than Sonnet 4.5 and Gemini-3-Flash, \textbf{suggesting credit table understanding depends more on visual perception and knowledge utilization than numerical calculation, which provides important guidance for further model optimization.}\\
\textbf{Table Visual Perception.} Results demonstrate that increasing table quantity degrades performance for most MLLMs. Interestingly, however, Gemini-3-Flash achieves its best performance in the [9,13] group. We attribute this phenomenon to the model's superior visual encoding capabilities: additional tables provide more visual evidence for reasoning rather than imposing computational burden, thereby enhancing inference efficiency and ultimately improving understanding performance. \textbf{This suggests that given sufficient visual capacity, increasing the number of input tables can actually benefit understanding by providing richer contextual information for reasoning.}\\
\textbf{Knowledge Utilization.} As shown in Figure \ref{tab:mllm_performance}, we observe that almost all MLLMs achieve the highest performance on Know.Percep and the lowest performance on Know.Reason. We attribute this phenomenon to the fact that directly retrieving required knowledge from tables through visual perception is considerably easier for MLLMs than inferring knowledge through contextual reasoning. This indicates that knowledge analysis and reasoning capabilities constitute the primary bottleneck for MLLMs in table understanding tasks. Additionally, for domain knowledge category, as demonstrated in Figure \ref{tab:mllm_performance_extended},we observe that different models exhibit varying sensitivity to different knowledge categories. Notably, Gemini-3-Flash shows lower sensitivity to temporal and ratio knowledge despite achieving the highest overall accuracy. \textbf{This suggests that knowledge capability optimization should adopt differentiated strategies tailored to each model's knowledge sensitivity profile.}
\section{Conclusion}
This work introduces MMFCTUB, a novel finance credit table understanding benchmark specifically designed to evaluate the capacities of MLLMs in credit table understanding. MMFCTUB encompass diverse practical credit tables and fine-grained capacities. Extensive evaluations on MMFCTUB allow us to identify significant performance limitations among existing MLLMs in table structure perception, domain knowledge utilization and numerical computation.
\section{Limitations}
Our evaluation focuses on the most frequently used table types in credit review processes. However, credit reports contain additional information that reflects applicants' economic profiles, and metrics computed solely from our selected tables may introduce bias in comprehensive credit assessment. Furthermore, while MMFCTUB incorporates diverse table types and quantities to measure MLLMs' table structure perception capabilities, finer-grained quantification would provide more interpretable insights. Specifically, operation-level granularity such as cross-row retrieval would offer more actionable guidance for improving MLLMs' true capacities in table understanding.

\clearpage
\bibliography{custom}

\clearpage
\clearpage
\clearpage
\newpage

\appendix

\section{Dataset Construct Prompts}
We introduced the dataset construction process in Section \ref{label_bench_contsuction}, which involves both advanced LLMs (GPT-5) and MLLMs (Gemini 2.5-Flash). To illustrate how these models interact with each pipeline component, we detail the prompts used in each generation stage. Given the overall user economic background distributions (average monthly income, credit score), table infrastructure specifications, and inter-table constraints, we employ GPT-5 to generate detailed loan accounts based on this information. As shown in Tables \ref{quasi_card_prompt_spec},\ref{general_loan_account_prompt}, and \ref{credit_card_prompt}, credit reports contain three account types: Quasi-Credit Card, General Loan Account, and Credit Card Account. Each account type includes common attributes (e.g., opening date, repayment records) as well as type-specific fields (e.g., overdraft balance for Credit Card Accounts). We therefore design specialized prompts for each account type. The LLM leverages its domain knowledge to understand the user's economic profile while adhering to account generation rules, producing account details that align with the user's background. All account fields are defined based on real-world credit report schemas. Notably, General Loan Accounts are further subdivided into three subtypes: Non-revolving Loan Account, Sub-account under Revolving Credit Limit, and Revolving Loan Account. Since these subtypes follow different table construction rules, we design subtype-specific prompts and generation procedures to ensure compliance with their respective structural constraints.

\section{Evaluation Prompts}
\begin{table*}[htbp]
\centering
\caption{Prompt for Quasi-Credit Card Account Data Generation}
\label{quasi_card_prompt_spec}
\small
\begin{tabularx}{\textwidth}{>{\raggedright\arraybackslash}p{3.5cm}X}
\toprule
\textbf{Section} & \textbf{Content} \\
\midrule
\textbf{Role Definition} & You are an experienced credit review expert with extensive experience in credit application approval. \\
\midrule
\textbf{Task Objective} & Generate the user's loan information based on the following requirements according to the user's credit status, economic situation, residential situation, and employment situation. \\
\midrule
\textbf{Input Parameters} & 
\begin{itemize}
    \item User's average monthly income: \{average\_month\_income\}
    \item User's credit rating: \{credit\_rating\}
    \item Residential situation: \{live\_tab\}
    \item Employment situation: \{prof\_tab\}
\end{itemize} \\
\midrule
\textbf{Output Requirement} & Generate this user's semi-credit card usage situation \\
\midrule
\textbf{Required Fields} & Issuing institution, Opening date, Account credit limit, Shared credit limit, Shared used limit, Currency, Business type, Guarantee method, Account status, Due date, Actual repayment this month, Cutoff date, Overdraft balance, Average overdraft balance in recent 6 months, Maximum overdraft balance, Unpaid balance overdue for more than 180 days, Total months of bad debt, Total months of overdue, Total months of settled, Repayment record start month, Repayment record end month, Account type, Total overdraft periods in repayment record, Current overdraft periods in repayment record \\
\midrule
\textbf{Account Status Options} & Normal, Overdue, Bad Debt, Settled \\
\midrule
\textbf{Constraint Rule} & (total months of bad debt + total months of overdue + total months of settled + n) $\leq$ m \newline
Where: \newline
$\bullet$ m = total months from repayment record start month to repayment record end month (inclusive) \newline
$\bullet$ n = random integer between 5-12 \\
\midrule
\textbf{Currency Assignment} & Value should be assigned based on user attributes \\
\midrule
\textbf{Field Format Examples} & 
\textbf{Date Fields:} Opening date: ``2020.05.15'' (YYYY.MM.DD); Cutoff date: ``2024.09.30''; Due date: ``2024.10.25''; Repayment record start month: ``2023-01'' (YYYY-MM); Repayment record end month: ``2023-12''. \textbf{Institutional Fields:} Issuing institution: ``Issuing Institution GQ''; Business type: ``Personal Semi-Credit Card''; Guarantee method: ``Credit Guarantee''; Account status: ``Normal''; Account type: ``Semi-Credit Card'' (fixed); Currency: ``RMB''. \textbf{Credit Limits:} Account credit limit: 50000; Shared credit limit: 50000; Shared used limit: 30000 ($\leq$ shared credit limit). \textbf{Balances:} Overdraft balance: 4500 ($<$ account credit limit); Average overdraft balance (6M): 3800; Maximum overdraft balance: 6200; Actual repayment this month: 3200; Unpaid balance overdue $>$180 days: 0. \textbf{Period Counts:} Total months of bad debt: 2; Total months of overdue: 3; Total months of settled: 2; Total overdraft periods: 7; Current overdraft periods: 5. \\
\midrule
\textbf{Output Format} & The account situation should be output in JSON object format. The overall output is a JSON object array, where each JSON object corresponds to one account. The keys within the object are the required fields for the account, and the values need to be reasonably assigned by you based on the user's specific situation. Note: Do not output any thinking process or extra content. Output strictly according to the specified format. \\
\bottomrule
\end{tabularx}
\end{table*}

\begin{table*}[htbp]
\centering
\caption{Prompt For General Loan Account Data Generation}
\label{general_loan_account_prompt}
\small
\begin{tabularx}{\textwidth}{>{\raggedright\arraybackslash}p{3.5cm}X}
\toprule
\textbf{Section} & \textbf{Content} \\
\midrule
\textbf{Role Definition} & You are an experienced credit review expert with extensive experience in credit application approval. \\
\midrule
\textbf{Task Objective} & You need to generate the user's loan information based on the following requirements according to the user's credit status, economic situation, residential situation, and employment situation. \\
\midrule
\textbf{Input Parameters} & 
\begin{itemize}
    \item User's average monthly income: \{average\_month\_income\}
    \item User's credit rating: \{credit\_rating\}
    \item Residential situation: \{live\_tab\}
    \item Employment situation: \{prof\_tab\}
\end{itemize} \\
\midrule
\textbf{Output Requirement} & Please generate this user's loan situation, including non-revolving loan accounts, sub-accounts under revolving credit limit, and revolving loan accounts \\
\midrule
\textbf{Required Fields} & Managing institution, Account credit limit, Currency, Business type, Guarantee method, Account status, Repayment periods, Remaining repayment periods, Actual repayment this month, Repayment method, Opening date, Closing date, Repayment record start month, Repayment record end month, Current overdue periods, Repayment due this month, Loan amount, Account type, Total months of bad debt, Total months of overdue, Total months of settled \\
\midrule
\textbf{Account Status Options} & Normal, Overdue, Bad Debt, Settled \\
\midrule
\textbf{Business Type Options} & Personal Commercial Housing Loan, Personal Housing Provident Fund Loan, Commercial Student Loan, Personal Consumer Loan, Cash Loan \\
\midrule
\textbf{Repayment Method Options} & Installment Equal Principal, Installment Equal Principal and Interest, One-time Principal and Interest at Maturity, Periodic Interest Settlement with Principal at Maturity, Periodic Interest Settlement with Flexible Principal Repayment, Equal Principal, Equal Principal and Interest \\
\midrule
\textbf{Account Type Options} & Non-revolving Loan Account, Sub-account under Revolving Credit Limit, Revolving Loan Account \\
\midrule
\textbf{Constraint Rule 1} & (total months of bad debt + total months of overdue + total months of settled + n) $\leq$ m \newline
Where: \newline
$\bullet$ m = total months from repayment record start month to repayment record end month (inclusive) \newline
$\bullet$ n = random integer between 5-12 \\
\midrule
\textbf{Constraint Rule 2} & Current overdue periods cannot exceed total months of overdue \\
\midrule
\textbf{Account Credit Limit Rule} & Only has value in revolving loan accounts, value should be assigned based on user attributes \\
\midrule
\textbf{Loan Amount Rule} & Only has value in non-revolving loan accounts and sub-accounts under revolving credit limit, value should be assigned based on user attributes \\
\midrule
\textbf{Repayment Periods Rule} & Empty for revolving loan accounts, a specific number for non-revolving loan accounts and sub-accounts under revolving credit limit, this number should be assigned based on user attributes \\
\midrule
\textbf{Current Overdue Periods Rule} & Value is determined based on account status, user attributes, and total months of overdue. Only when account status is Overdue, its value is greater than 0, otherwise it equals 0 \\
\midrule
\textbf{Currency Assignment} & Value should be assigned based on user attributes \\
\midrule
\textbf{Field Format Examples} & 
\textbf{Date Fields:} opening\_date=``2013.07.22'' (YYYY.MM.DD); closing\_date=``2015-05-05'' (YYYY-MM-DD); repayment record start month=``2023-01'' (YYYY-MM); repayment record end month=``2023-12''. \textbf{Account Type:} account\_type=``Revolving Loan Account'' (options: Non-revolving Loan Account, Sub-account under Revolving Credit Limit, Revolving Loan Account). \textbf{Business \& Methods:} business\_type=``Personal Consumer Loan'' (options: Personal Commercial Housing Loan, Personal Housing Provident Fund Loan, Commercial Student Loan, Stock Pledge Repo Transaction, Personal Consumer Loan); ... \\
\midrule
\textbf{Output Format} & The account situation should be output in JSON object format. The overall output is a JSON object array, where each JSON object corresponds to one account. The keys within the object are the required fields for the account, and the values need to be reasonably assigned by you based on the user's specific situation. Note: Do not output any thinking process or extra content. Output strictly according to the specified format. \\
\bottomrule
\end{tabularx}
\end{table*}

\begin{table*}[htbp]
\centering
\caption{Prompt Credit Card Account Data Generation }
\label{credit_card_prompt}
\small
\begin{tabularx}{\textwidth}{>{\raggedright\arraybackslash}p{3.5cm}X}
\toprule
\textbf{Section} & \textbf{Content} \\
\midrule
\textbf{Role Definition} & You are an experienced credit review expert with extensive experience in credit application approval. \\
\midrule
\textbf{Task Objective} & You need to generate the user's loan information based on the following requirements according to the user's credit status, economic situation, residential situation, and employment situation. \\
\midrule
\textbf{Input Parameters} & 
\begin{itemize}
    \item User's average monthly income: \{average\_month\_income\}
    \item User's credit rating: \{credit\_rating\}
    \item Residential situation: \{live\_tab\}
    \item Employment situation: \{prof\_tab\}
\end{itemize} \\
\midrule
\textbf{Output Requirement} & Please generate this user's credit card usage situation \\
\midrule
\textbf{Required Fields} & Issuing institution, Shared credit limit, Shared used limit, Used limit, Average used limit in recent 6 months, Maximum used limit, Remaining installment periods, Opening date, Cutoff date, Repayment record start month, Repayment record end month, Account credit limit, Currency, Business type, Guarantee method, Account status, Due date, Repayment due this month, Actual repayment this month, Current overdue periods, Total periods of bad debt, Total months of overdue, Total months of settled, Business type \\
\midrule
\textbf{Account Status Options} & Normal, Overdue, Bad Debt, Settled \\
\midrule
\textbf{Constraint Rule 1} & (total months of bad debt + total months of overdue + total months of settled + n) $\leq$ m \newline
Where: \newline
$\bullet$ m = total months from repayment record start month to repayment record end month (inclusive) \newline
$\bullet$ n = random integer between 5-12 \\
\midrule
\textbf{Constraint Rule 2} & Current overdue periods cannot exceed total months of overdue \\
\midrule
\textbf{Current Overdue Periods Rule} & Value is determined based on account status, user attributes, and total months of overdue. Only when account status is Overdue, its value is greater than 0, otherwise it equals 0 \\
\midrule
\textbf{Currency Assignment} & Value should be assigned based on user attributes \\
\midrule
\textbf{Field Format Examples} & 
\textbf{Date Fields:} opening\_date=``2020.05.10'' (YYYY.MM.DD); cutoff\_date=``2024.10.31''; due\_date=``2024.10.25''; start\_time=``2023-01'' (YYYY-MM); end\_time=``2023-12''. \textbf{Account Type:} account\_type=``Credit Card'' (fixed, can only be Credit Card). \textbf{Status \& Institution:} issuing\_institution=``Issuing Institution AQ''; business\_type=``Credit Card''; guarantee\_method=``Unsecured''; account\_status=``Overdue''; currency=``RMB''. \textbf{Credit Limits:} shared\_credit\_limit=``50000''; shared\_used\_limit=``30000'' ($\leq$ shared credit limit); credit\_limit=``50000'' (independent); used\_limit=``32000'' (independent usage); avg\_used\_limit\_6m=``28000''; max\_used\_limit=``45000''. \textbf{Repayment Fields:} this\_month\_due=``5000'' (estimated based on status, balance, method, overdue periods); actual\_repayment=``0'' (when status is Overdue or Bad Debt: actual $<$ due; otherwise: actual $=$ due); remaining\_installment\_periods=``12''. \textbf{Period Counts:} b\_count=2 (total months of bad debt); num\_count=3 (total months of overdue); c\_count=2 (total months of settled); current\_overdue\_periods=``3'' (must be $\leq$ num\_count, $>$0 only when status is Overdue). \\
\midrule
\textbf{Output Format} & The account situation should be output in JSON object format. The overall output is a JSON object array, where each JSON object corresponds to one account. The keys within the object are the required fields for the account, and the values need to be reasonably assigned by you based on the user's specific situation. Note: Do not output any thinking process or extra content. Output strictly according to the specified format. \\
\bottomrule
\end{tabularx}
\end{table*}

\begin{figure*}[!htbp]  
    \centering
    \includegraphics[width=\textwidth]{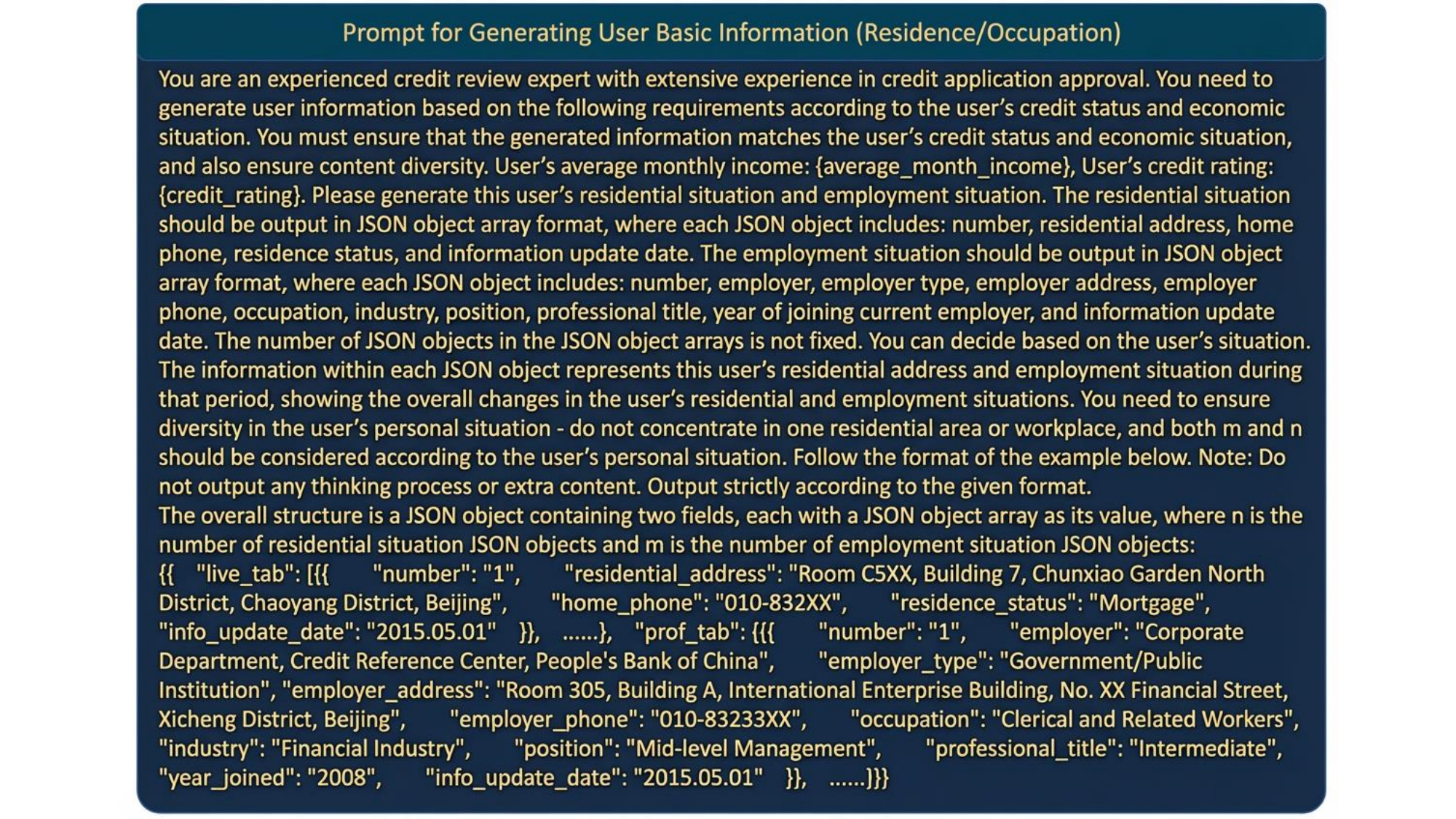}  
    \captionsetup{justification=centering, singlelinecheck=false}
    \caption{Prompt for User Basic Information (Residence/Occupation).}  
    \label{prompts4table_user_resid_prof} 
\end{figure*}
\begin{figure*}[!htbp]  
    \centering
    \includegraphics[width=\textwidth]{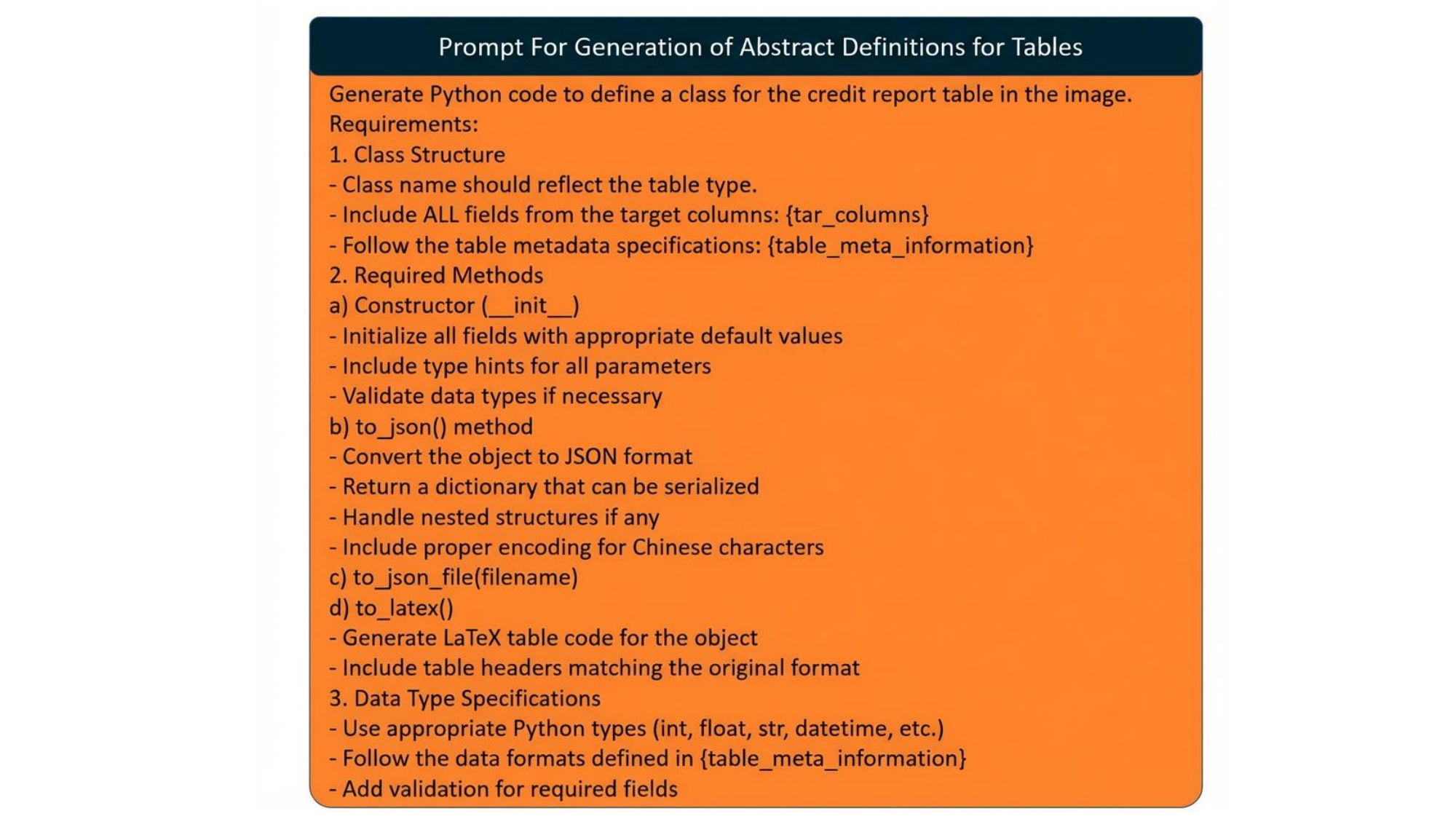}  
    \captionsetup{justification=centering, singlelinecheck=false}
    \caption{Prompt for Generation of Abstract Definitions for Tables.}  
    \label{prompts4table_abs_def} 
\end{figure*}
\begin{figure*}[!htbp]  
    \centering
    \includegraphics[width=\textwidth]{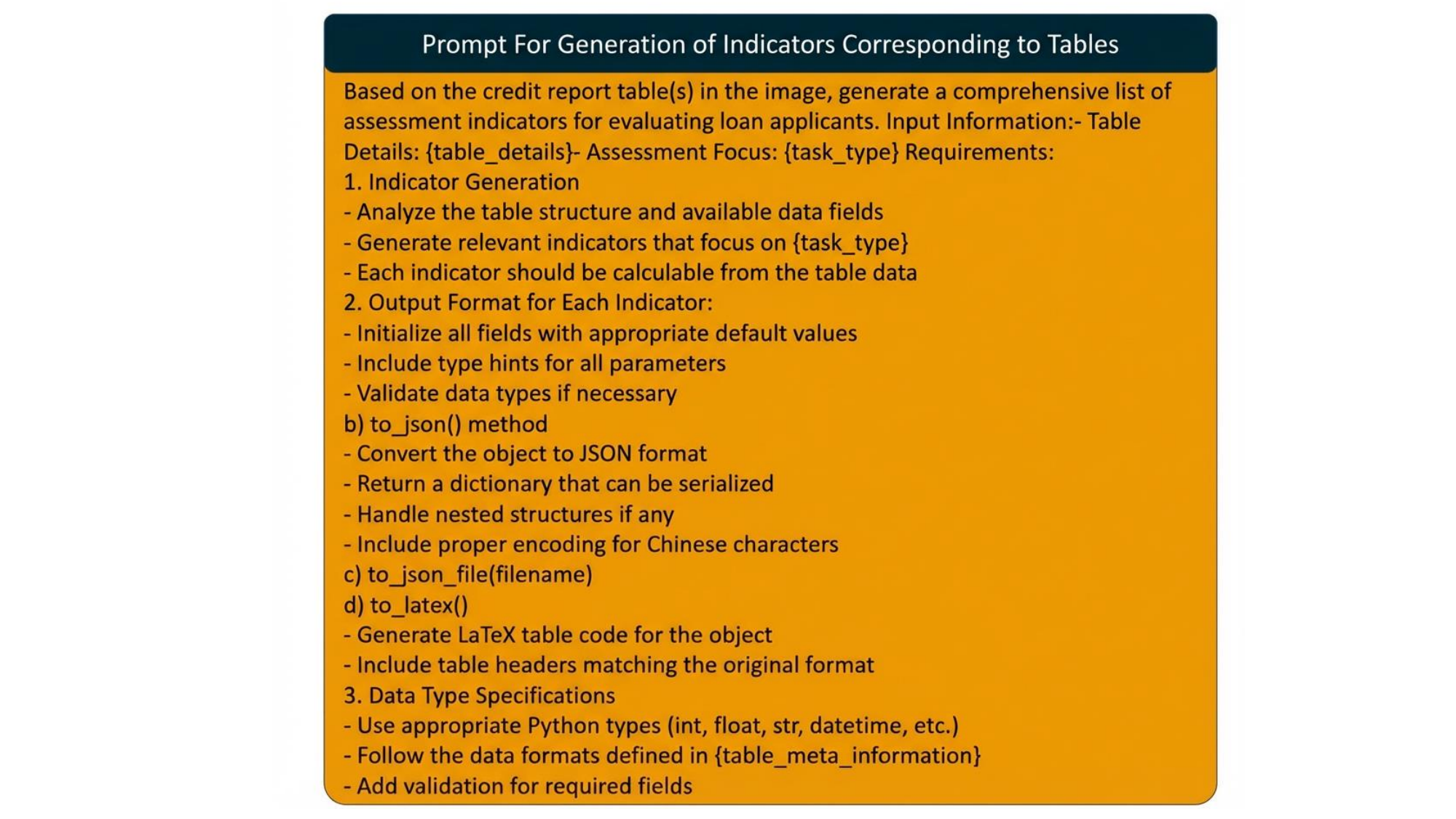}  
    \captionsetup{justification=centering, singlelinecheck=false}
    \caption{Prompt for Generation of Indicators Corresponding to Tables.}  
    \label{prompts4table_indicators} 
\end{figure*}
\begin{figure*}[!htbp]  
    \centering
    \includegraphics[width=\textwidth]{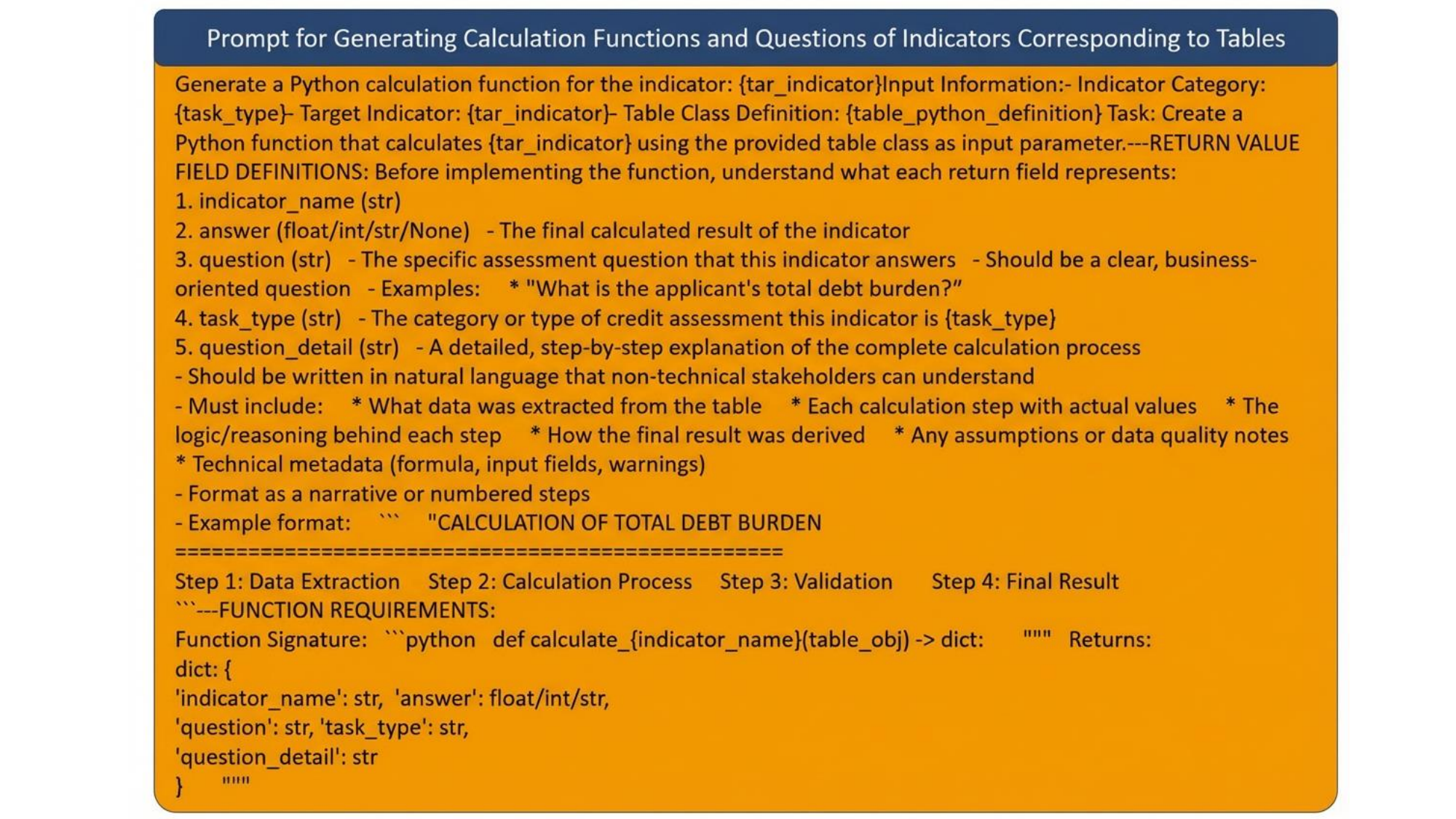}  
    \captionsetup{justification=centering, singlelinecheck=false}
    \caption{Prompt for Generating Calculation Functions and Questions of Indicators Corresponding to Tables.}  
    \label{prompts4table_ques_func} 
\end{figure*}
\begin{figure*}[!htbp]  
    \centering
    \includegraphics[width=\textwidth]{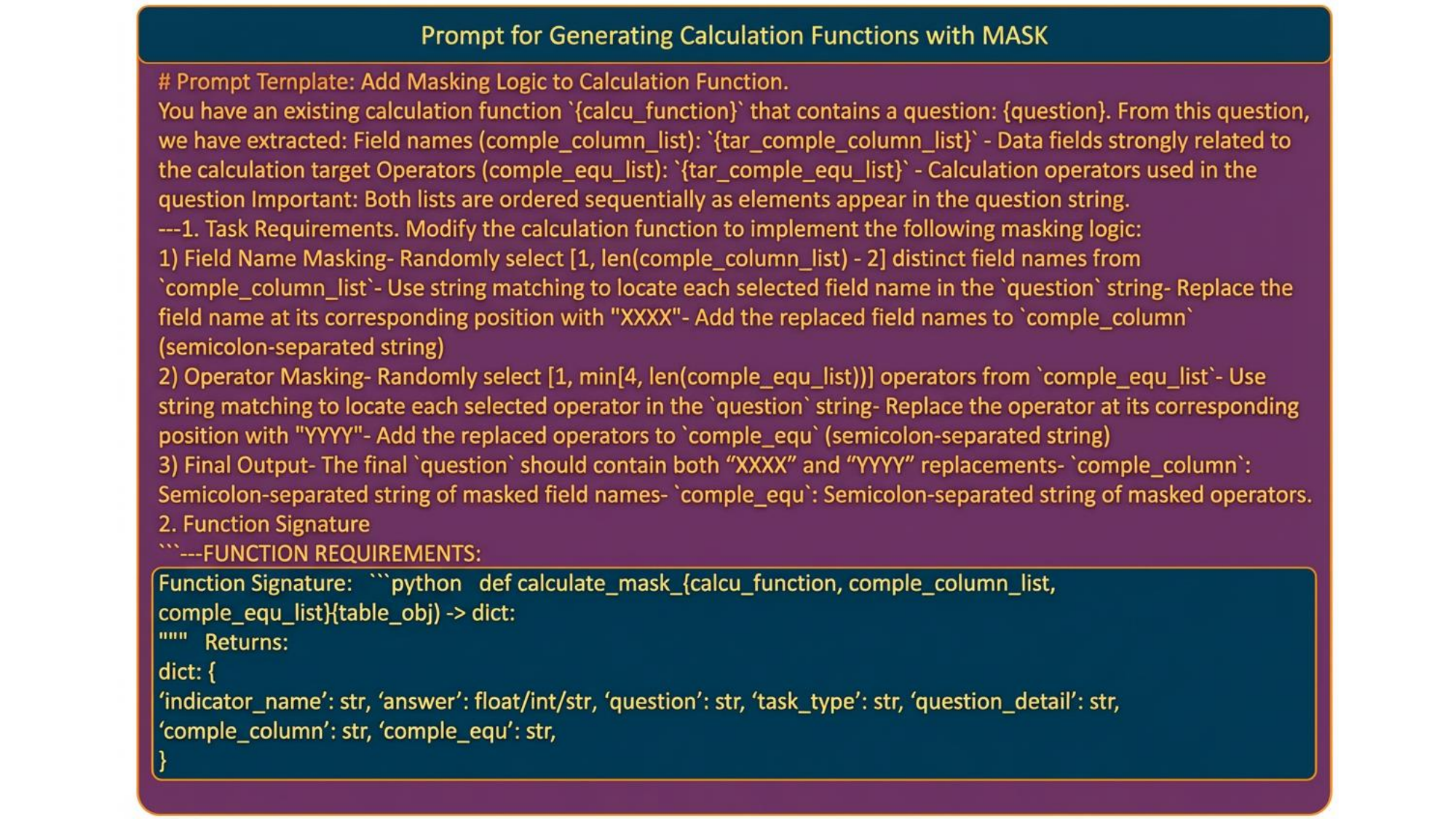}  
    \captionsetup{justification=centering, singlelinecheck=false}
    \caption{Prompt for Generating Calculation Functions with MASK.}  
    \label{prompts4table_ques_func_mask} 
\end{figure*}

\begin{figure*}[!htbp]  
    \centering
    \includegraphics[width=\textwidth]{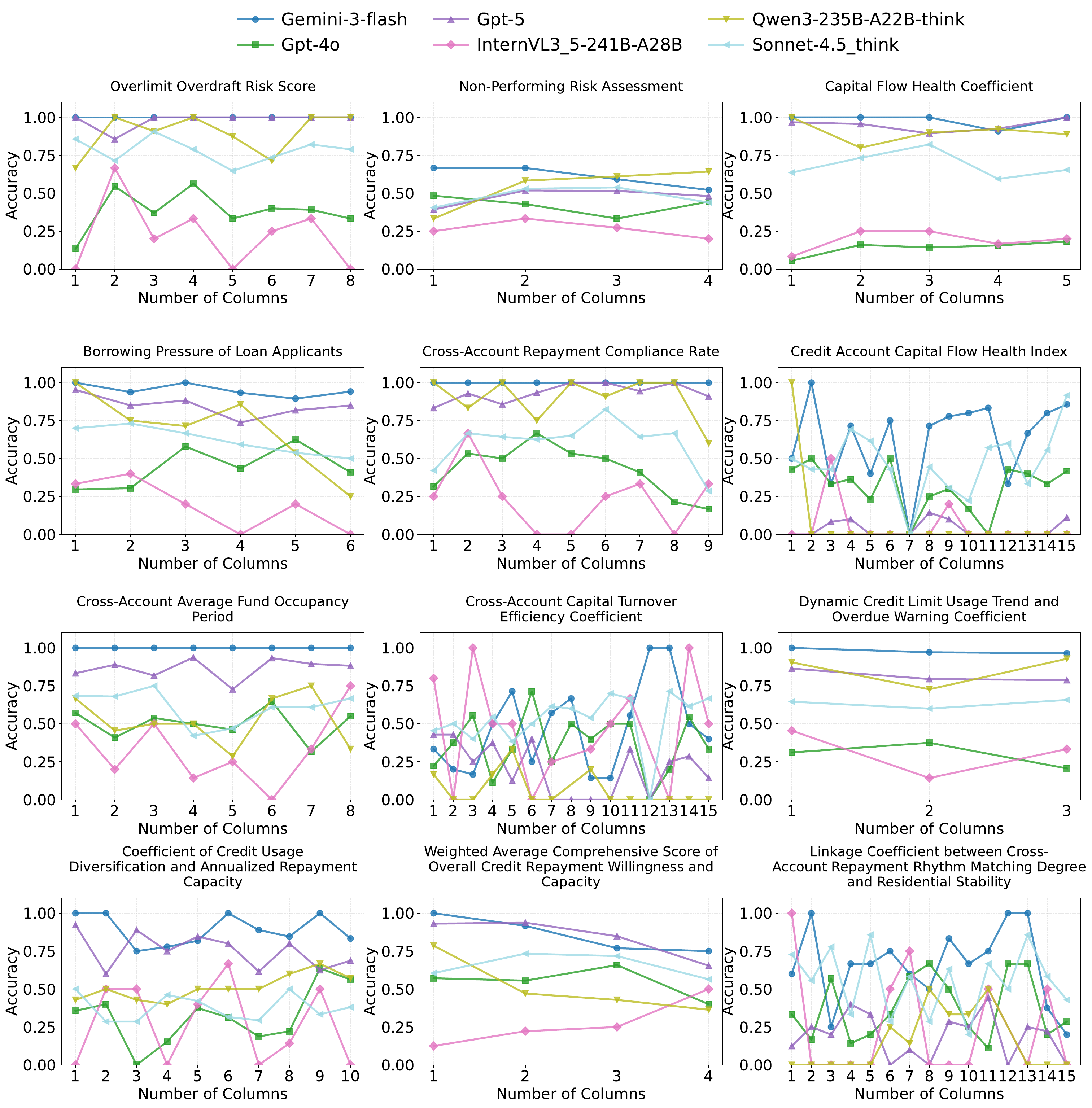}  
    \captionsetup{justification=centering, singlelinecheck=false}
    \caption{Relationship between the number of knowledge elements to be supplemented for a single metric (denoted as Number of Columns in the prompt) and the final table understanding (TU) accuracy. Here, Number of Columns refers to the quantity of knowledge elements that need to be recovered in the prompt, while Accuracy denotes the resulting TU accuracy across different metrics.}  
    \label{metric_category_accuracy_vs_column_num_selection} 
\end{figure*}

\begin{figure*}[!htbp]  
    \centering
    \includegraphics[width=\textwidth]{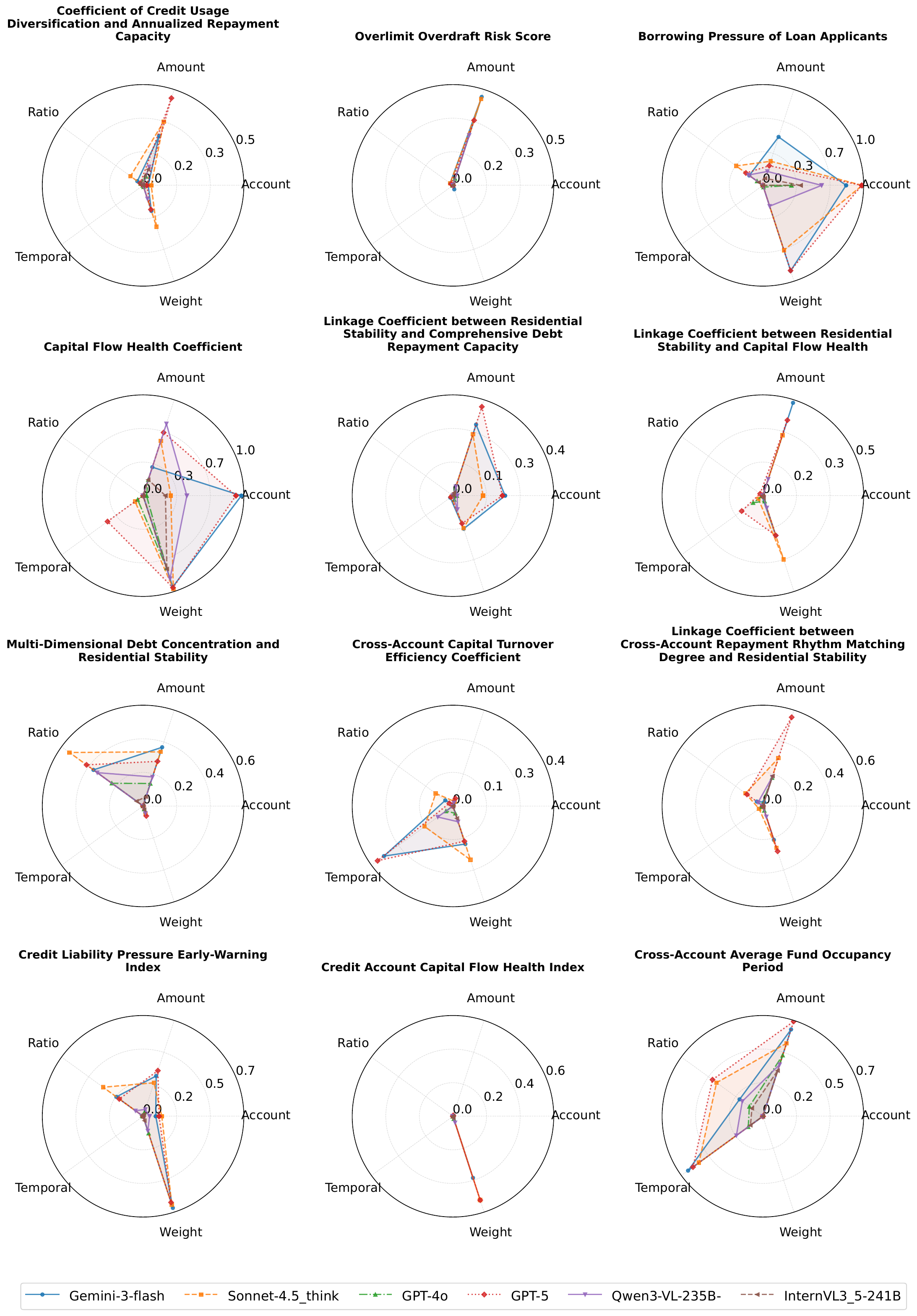}  
    \captionsetup{justification=centering, singlelinecheck=false}
    \caption{Demonstrate the distribution of knowledge hit rates across different metrics.}  
    \label{radar_all_metrics} 
\end{figure*}

\begin{figure*}[!htbp]  
    \centering
    \includegraphics[width=\textwidth]{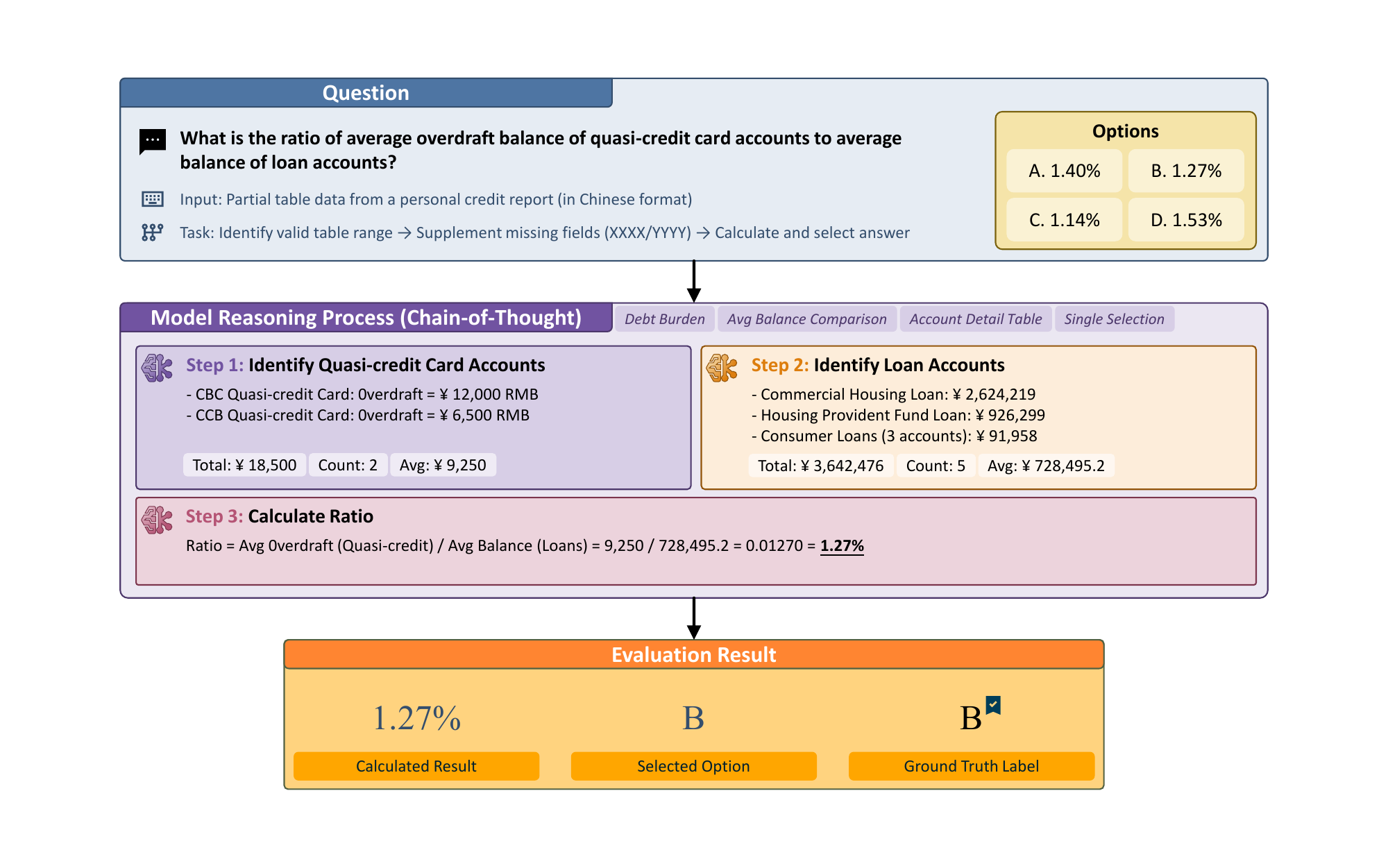}  
    \captionsetup{justification=centering, singlelinecheck=false}
    \caption{Demonstrate of Table Structure Perception}  
    \label{fintab_case_1} 
\end{figure*}

\begin{figure*}[!htbp]  
    \centering
    \includegraphics[width=\textwidth]{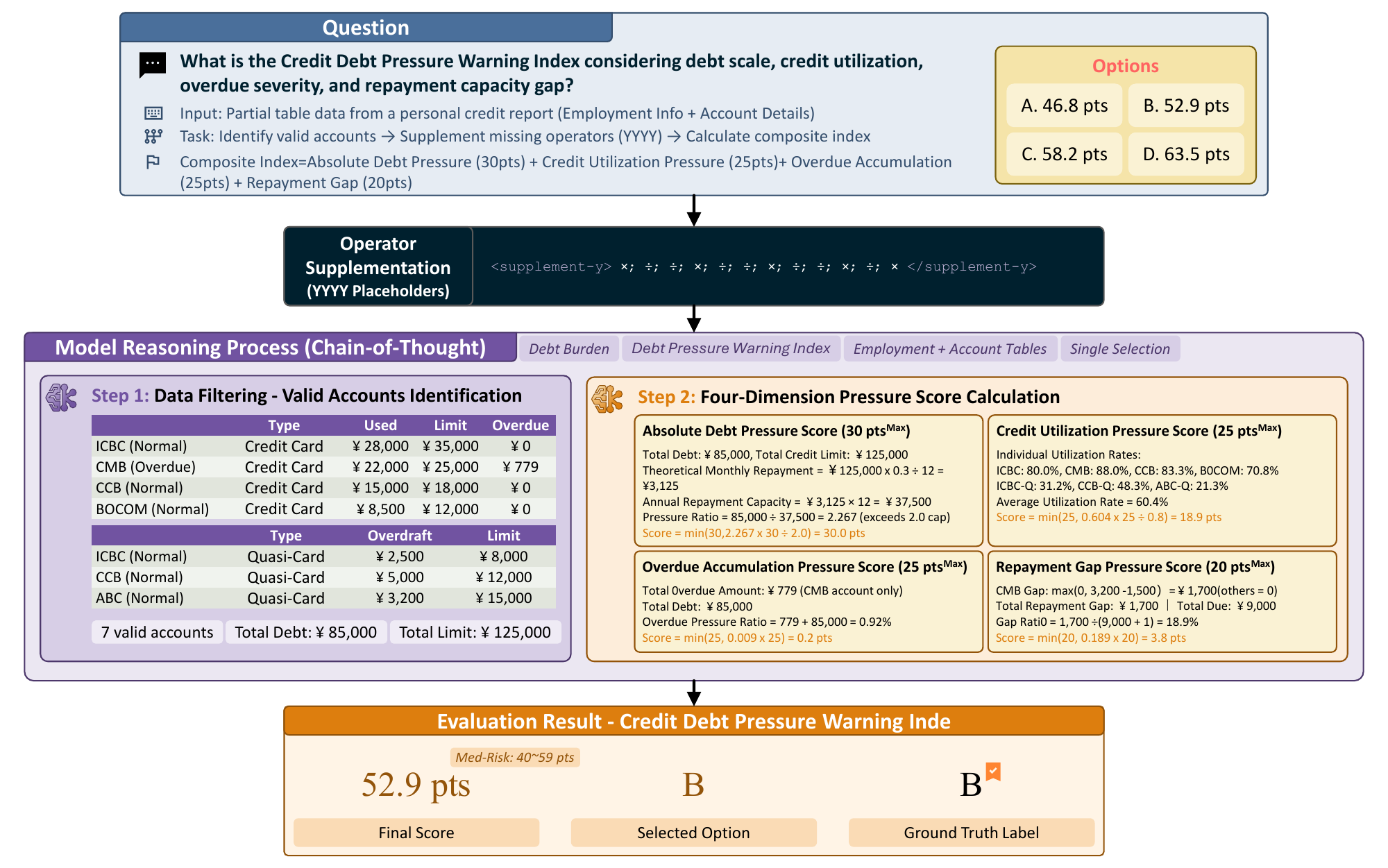}  
    \captionsetup{justification=centering, singlelinecheck=false}
    \caption{Demonstrate of Domain Knowledge Utilization}  
    \label{fintab_case_2} 
\end{figure*}

\begin{figure*}[!htbp]  
    \centering
    \includegraphics[width=\textwidth]{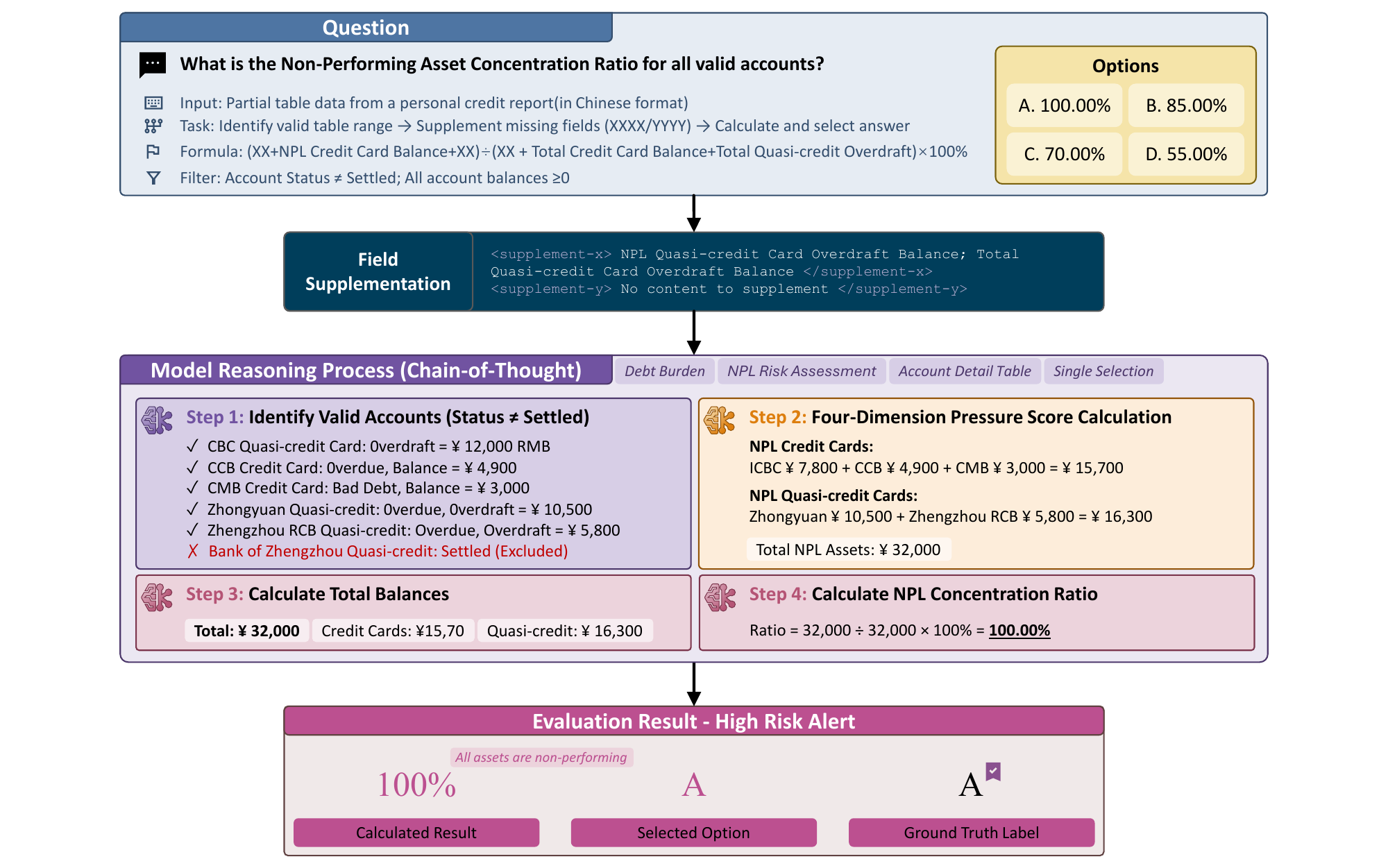}  
    \captionsetup{justification=centering, singlelinecheck=false}
    \caption{Demonstrate of Numerical Computation}  
    \label{fintab_case_3} 
\end{figure*}

\section{Expert Consultation Process Record}

The design and validation of our research framework benefited significantly from consultations with senior professionals at one of the largest financial institutions, which serves over 2 billion users globally. Through structured interviews with industry specialists, we gathered critical insights into consumer credit evaluation. These expert perspectives directly shaped the development of  this benchmark. Our consultation panel comprised seasoned specialists across credit underwriting, risk control, and lending operations, each with 8 to over 15 years of hands-on experience. Each expert brings unique domain knowledge spanning different aspects of the credit evaluation lifecycle, from initial application review to final credit decision-making.

Key findings from our expert consultations highlighted several critical aspects:

\textbf{Expert 1 (14 years, Senior Credit Underwriter)} Specializes in consumer credit assessment and loan approval decision-making. Extensive experience in interpreting multi-table credit data including credit reports, bank statements, and asset verification documents.

\textbf{Expert 2 (12 years, Credit Risk Manager)} Focuses on credit risk modeling and portfolio risk management. Expert in analyzing credit bureau reports, identifying default indicators across multiple data tables, and developing risk assessment frameworks.

\textbf{Expert 3 (11 years, Credit Review Specialist)} Concentrates on credit application review and financial document verification. Proficient in cross-referencing information across credit tables, detecting inconsistencies, and validating borrower credentials.

\textbf{Expert 4 (9 years, Lending Operations Manager)} Manages end-to-end credit approval workflows and automation systems. Deep understanding of credit table processing requirements, data integration challenges, and operational efficiency optimization.

\textbf{Expert 5 (10 years, Credit Policy Analyst)} Designs and implements credit underwriting policies and scoring models. Expertise in credit scoring methodology, regulatory compliance (Basel III, local lending regulations), and credit table standardization.

\textbf{Expert 6 (8 years, Credit Data Analyst)} Specializes in quantitative credit analysis and tabular data processing. Expert in extracting risk signals from structured credit data, performing multi-table joint analysis, and building predictive models for creditworthiness assessment.

The expert panel provided invaluable guidance on three critical aspects. First, the practical challenges in processing and integrating multi-table credit data, including credit reports and asset statements etc.; Second, the information dependencies and cross-validation requirements across different credit tables that underwriters must consider. Lastly, the real-world task requirements for credit assessment, including question answering about borrower profiles, fact verification across multiple data sources, and comprehensive creditworthiness summarization. Their domain expertise ensured that our benchmark addresses authentic industry needs and accurately reflects the complexity of actual credit evaluation workflows in contemporary consumer lending operations.

\subsection{Dataset Details}

\subsection{Question Sample} 

This benchmark evaluates the model's capabilities in processing credit report tables for credit assessment tasks across three dimensions: table structure perception, domain knowledge application, and numerical computation. The result of each metric serves as part of the basis for the final credit assessment decision. These metrics primarily evaluate loan applicants from four perspectives: repayment capacity, repayment willingness, debt pressure, and cash flow conditions.

\begin{table}[p]
\centering
\renewcommand\arraystretch{1.5}
\begin{minipage}{\textwidth}
\centering
\resizebox{\textwidth}{!}{
\begin{tabular}{|c|c|c|c|c|c|c|c|c|c|c|c|c|}
\hline
\multicolumn{13}{|c|}{\textbf{Account (Credit Agreement ID: ZDJK20250015)}} \\
\hline
\multicolumn{3}{|c|}{Managing Institution} & \multicolumn{3}{|c|}{Opening Date} & \multicolumn{3}{|c|}{Account Credit Limit} & \multicolumn{4}{|c|}{Loan Amount} \\
\hline
\multicolumn{3}{|c|}{Commercial Bank PD} & \multicolumn{3}{|c|}{2020.08.15} & \multicolumn{3}{|c|}{2849450} & \multicolumn{4}{|c|}{2800000} \\
\hline
\multicolumn{3}{|c|}{Account Type} & \multicolumn{3}{|c|}{Currency} & \multicolumn{3}{|c|}{Business Type} & \multicolumn{4}{|c|}{Guarantee Type} \\
\hline
\multicolumn{3}{|c|}{Non-revolving Loan Account} & \multicolumn{3}{|c|}{CNY} & \multicolumn{3}{|c|}{Personal Commercial Housing Loan} & \multicolumn{4}{|c|}{Mortgage} \\
\hline
\multicolumn{3}{|c|}{Account Status} & \multicolumn{3}{|c|}{Five-tier Classification} & \multicolumn{3}{|c|}{Payment Due Date} & \multicolumn{4}{|c|}{Monthly Payment Due} \\
\hline
\multicolumn{3}{|c|}{Normal} & \multicolumn{3}{|c|}{N/A} & \multicolumn{3}{|c|}{2023.12.25} & \multicolumn{4}{|c|}{15800} \\
\hline
\multicolumn{3}{|c|}{Actual Monthly Payment} & \multicolumn{3}{|c|}{Last Payment Date} & \multicolumn{3}{|c|}{Total Payment Terms} & \multicolumn{4}{|c|}{Remaining Payment Terms} \\
\hline
\multicolumn{3}{|c|}{15800} & \multicolumn{3}{|c|}{2023.12.25} & \multicolumn{3}{|c|}{360} & \multicolumn{4}{|c|}{318} \\
\hline
\multicolumn{3}{|c|}{Maturity Date} & \multicolumn{3}{|c|}{Current Overdue Periods} & \multicolumn{3}{|c|}{Current Overdue Amount} & \multicolumn{4}{|c|}{As of Date} \\
\hline
\multicolumn{3}{|c|}{2050.07.31} & \multicolumn{3}{|c|}{0} & \multicolumn{3}{|c|}{0} & \multicolumn{4}{|c|}{Aug 15, 2050} \\
\hline
\multicolumn{6}{|c|}{Principal Overdue 31-60 Days} & \multicolumn{7}{|c|}{Principal Overdue 61-90 Days} \\
\hline
\multicolumn{6}{|c|}{0} & \multicolumn{7}{|c|}{0} \\
\hline
\multicolumn{13}{|c|}{Balance: 2624219} \\
\hline
\multicolumn{13}{|c|}{Bad Debt Periods: N/A} \\
\hline
\multicolumn{13}{|c|}{\textbf{Repayment History | As of 2023-12}} \\
\hline
& 1 & 2 & 3 & 4 & 5 & 6 & 7 & 8 & 9 & 10 & 11 & 12 \\
\hline
\multirow{2}{*}{2023} & N & N & N & N & N & N & N & N & N & N & N & N \\
\cline{2-13}
& 0.00 & 0.00 & 0.00 & 0.00 & 0.00 & 0.00 & 0.00 & 0.00 & 0.00 & 0.00 & 0.00 & 0.00 \\
\hline
\multirow{2}{*}{2022} & N & N & N & N & N & N & N & N & N & N & N & N \\
\cline{2-13}
& 0.00 & 0.00 & 0.00 & 0.00 & 0.00 & 0.00 & 0.00 & 0.00 & 0.00 & 0.00 & 0.00 & 0.00 \\
\hline
\multirow{2}{*}{2021} & N & N & N & N & N & N & N & N & N & N & N & N \\
\cline{2-13}
& 0.00 & 0.00 & 0.00 & 0.00 & 0.00 & 0.00 & 0.00 & 0.00 & 0.00 & 0.00 & 0.00 & 0.00 \\
\hline
\multirow{2}{*}{2020} & N/A & N/A & N/A & N/A & N/A & N/A & N/A & N/A & N & N & N & N \\
\cline{2-13}
& N/A & N/A & N/A & N/A & N/A & N/A & N/A & N/A & 0.00 & 0.00 & 0.00 & 0.00 \\
\hline
\multicolumn{13}{|c|}{Account Closure Date: N/A} \\
\hline
\end{tabular}
}
\captionof{table}{Credit Account Table Details}
\label{tab:credit_account}
\end{minipage}
\end{table}
\clearpage

\begin{table}[p]
\centering
\renewcommand\arraystretch{1.5}
\begin{minipage}{\textwidth}
\centering
\resizebox{\textwidth}{!}{
\begin{tabular}{|C{2cm}|C{1cm}|C{1cm}|C{1cm}|C{1cm}|C{1cm}|C{1cm}|C{1cm}|C{1cm}|C{1cm}|C{1cm}|C{1cm}|C{1cm}|}
\hline
\multicolumn{13}{|c|}{\textbf{Account (Credit Agreement ID: ZDJK20250015)}} \\
\hline
\multicolumn{3}{|C{3.5cm}|}{Managing Institution} & \multicolumn{3}{C{3.5cm}|}{Opening Date} & \multicolumn{3}{C{3.5cm}|}{Account Credit Limit} & \multicolumn{4}{C{4cm}|}{Loan Amount} \\
\hline
\multicolumn{3}{|C{3.5cm}|}{Commercial Bank XH} & \multicolumn{3}{C{3.5cm}|}{2022.05.20} & \multicolumn{3}{C{3.5cm}|}{279066} & \multicolumn{4}{C{4cm}|}{180000} \\
\hline
\multicolumn{3}{|C{3.5cm}|}{Account Type} & \multicolumn{3}{C{3.5cm}|}{Currency} & \multicolumn{3}{C{3.5cm}|}{Business Type} & \multicolumn{4}{C{4cm}|}{Guarantee Type} \\
\hline
\multicolumn{3}{|C{3.5cm}|}{Sub-account under Revolving Credit} & \multicolumn{3}{C{3.5cm}|}{CNY} & \multicolumn{3}{C{3.5cm}|}{Personal Consumer Loan} & \multicolumn{4}{C{4cm}|}{Unsecured} \\
\hline
\multicolumn{3}{|C{3.5cm}|}{Account Status} & \multicolumn{3}{C{3.5cm}|}{Five-tier Classification} & \multicolumn{3}{C{3.5cm}|}{Payment Due Date} & \multicolumn{4}{C{4cm}|}{Monthly Payment Due} \\
\hline
\multicolumn{3}{|C{3.5cm}|}{Normal} & \multicolumn{3}{C{3.5cm}|}{N/A} & \multicolumn{3}{C{3.5cm}|}{2023.12.18} & \multicolumn{4}{C{4cm}|}{5580} \\
\hline
\multicolumn{3}{|C{3.5cm}|}{Actual Monthly Payment} & \multicolumn{3}{C{3.5cm}|}{Last Payment Date} & \multicolumn{3}{C{3.5cm}|}{Total Payment Terms} & \multicolumn{4}{C{4cm}|}{Remaining Payment Terms} \\
\hline
\multicolumn{3}{|C{3.5cm}|}{5580} & \multicolumn{3}{C{3.5cm}|}{2023.12.18} & \multicolumn{3}{C{3.5cm}|}{36} & \multicolumn{4}{C{4cm}|}{16} \\
\hline
\multicolumn{3}{|C{3.5cm}|}{Maturity Date} & \multicolumn{3}{C{3.5cm}|}{Current Overdue Periods} & \multicolumn{3}{C{3.5cm}|}{Current Overdue Amount} & \multicolumn{4}{C{4cm}|}{As of Date} \\
\hline
\multicolumn{3}{|C{3.5cm}|}{2025.04.30} & \multicolumn{3}{C{3.5cm}|}{0} & \multicolumn{3}{C{3.5cm}|}{0} & \multicolumn{4}{C{4cm}|}{May 20, 2025} \\
\hline
\multicolumn{6}{|C{6cm}|}{Principal Overdue 31-60 Days} & \multicolumn{7}{C{7cm}|}{Principal Overdue 61-90 Days} \\
\hline
\multicolumn{6}{|C{6cm}|}{0} & \multicolumn{7}{C{7cm}|}{0} \\
\hline
\multicolumn{13}{|C{13cm}|}{Balance: 91958} \\
\hline
\multicolumn{13}{|c|}{Bad Debt Periods: N/A} \\
\hline
\multicolumn{13}{|c|}{\textbf{Repayment History | As of 2023-12}} \\
\hline
& 1 & 2 & 3 & 4 & 5 & 6 & 7 & 8 & 9 & 10 & 11 & 12 \\
\hline

\multicolumn{1}{|C{1cm}|}{\multirow{2}{*}{2023}} & N & N & N & N & N & N & N & N & N & N & N & N \\
\cline{2-13}
\multicolumn{1}{|C{1cm}|}{} & 0.00 & 0.00 & 0.00 & 0.00 & 0.00 & 0.00 & 0.00 & 0.00 & 0.00 & 0.00 & 0.00 & 0.00 \\
\hline

\multicolumn{1}{|C{1cm}|}{\multirow{2}{*}{2022}} & N/A & N/A & N/A & N/A & N/A & N & N & N & N & N & N & N \\
\cline{2-13}
\multicolumn{1}{|C{1cm}|}{} & N/A & N/A & N/A & N/A & N/A & 0.00 & 0.00 & 0.00 & 0.00 & 0.00 & 0.00 & 0.00 \\
\hline

\multicolumn{13}{|c|}{Account Closure Date: N/A} \\
\hline
\end{tabular}
}
\captionof{table}{Credit Account Details}
\label{tab:credit_account_revolving}
\end{minipage}
\end{table}

\begin{table}[p]
\centering
\renewcommand\arraystretch{1.2}
\begin{minipage}{\textwidth}
\centering
\begin{tabular}{|c|p{4cm}|c|p{4cm}|}
\hline
\multicolumn{4}{|c|}{\textbf{Occupation Information (ID: 1)}} \\
\hline
ID & 1 & Company & Zhengzhou Shunda Logistics Co., Ltd. \\
\hline
Company Nature & Private Enterprise & Company Address & No. 12, Area B, Logistics Park, Intersection of West 3rd Ring Road and Longhai Road, Zhongyuan District, Zhengzhou City, Henan Province \\
\hline
Company Phone & 0371—67123456 & Occupation & Commerce and Service Personnel \\
\hline
Industry & Transportation, Storage and Postal Services & Position & General Staff \\
\hline
Title & None & Entry Year & 2021 \\
\hline
Update Date & 2023.08.15 &  &  \\
\hline
\end{tabular}
\captionof{table}{Occupation Information Table Details}
\label{tab:occupation_info}
\end{minipage}
\end{table}
\clearpage

\begin{table}[p]
\centering
\renewcommand\arraystretch{1.5}
\begin{minipage}{\textwidth}
\centering
\resizebox{\textwidth}{!}{
\begin{tabular}{|c|c|c|c|c|c|c|c|c|c|c|c|c|}
\hline
\multicolumn{13}{|c|}{\textbf{Account (Credit Agreement ID: ZDJK20250015)}} \\
\hline
\multicolumn{3}{|c|}{Managing Institution} & \multicolumn{3}{c|}{Opening Date} & \multicolumn{3}{c|}{Account Credit Limit} & \multicolumn{4}{c|}{Loan Amount} \\
\hline
\multicolumn{3}{|c|}{Commercial Bank ZJ} & \multicolumn{3}{c|}{2021.11.10} & \multicolumn{3}{c|}{300000} & \multicolumn{4}{c|}{238907} \\
\hline
\multicolumn{3}{|c|}{Account Type} & \multicolumn{3}{c|}{Currency} & \multicolumn{3}{c|}{Business Type} & \multicolumn{4}{c|}{Guarantee Type} \\
\hline
\multicolumn{3}{|c|}{Revolving Loan Account} & \multicolumn{3}{c|}{CNY} & \multicolumn{3}{c|}{Personal Consumer Loan} & \multicolumn{4}{c|}{Unsecured} \\
\hline
\multicolumn{3}{|c|}{Account Status} & \multicolumn{3}{c|}{Five-tier Classification} & \multicolumn{3}{c|}{Payment Due Date} & \multicolumn{4}{c|}{Monthly Payment Due} \\
\hline
\multicolumn{3}{|c|}{Normal} & \multicolumn{3}{c|}{N/A} & \multicolumn{3}{c|}{2023.12.21} & \multicolumn{4}{c|}{1200} \\
\hline
\multicolumn{3}{|c|}{Actual Monthly Payment} & \multicolumn{3}{c|}{Last Payment Date} & \multicolumn{3}{c|}{Total Payment Terms} & \multicolumn{4}{c|}{Remaining Payment Terms} \\
\hline
\multicolumn{3}{|c|}{1200} & \multicolumn{3}{c|}{2023.12.21} & \multicolumn{3}{c|}{N/A} & \multicolumn{4}{c|}{N/A} \\
\hline
\multicolumn{3}{|c|}{Maturity Date} & \multicolumn{3}{c|}{Current Overdue Periods} & \multicolumn{3}{c|}{Current Overdue Amount} & \multicolumn{4}{c|}{As of Date} \\
\hline
\multicolumn{3}{|c|}{2024.03.21} & \multicolumn{3}{c|}{0} & \multicolumn{3}{c|}{0} & \multicolumn{4}{c|}{Nov 10, 2026} \\
\hline
\multicolumn{6}{|c|}{Principal Overdue 31-60 Days} & \multicolumn{7}{c|}{Principal Overdue 61-90 Days} \\
\hline
\multicolumn{6}{|c|}{0} & \multicolumn{7}{c|}{0} \\
\hline
\multicolumn{13}{|c|}{Balance: 0} \\
\hline
\multicolumn{13}{|c|}{Bad Debt Periods: N/A} \\
\hline
\multicolumn{13}{|c|}{\textbf{Repayment History | As of 2023-12}} \\
\hline
& 1 & 2 & 3 & 4 & 5 & 6 & 7 & 8 & 9 & 10 & 11 & 12 \\
\hline

\multicolumn{1}{|c|}{\multirow{2}{*}{2023}} & N & N & N & N & N & N & N & N & N & N & N & N \\
\cline{2-13}
\multicolumn{1}{|c|}{} & 0.00 & 0.00 & 0.00 & 0.00 & 0.00 & 0.00 & 0.00 & 0.00 & 0.00 & 0.00 & 0.00 & 0.00 \\
\hline

\multicolumn{1}{|c|}{\multirow{2}{*}{2022}} & N & N & N & N & N & N & N & N & N & N & N & N \\
\cline{2-13}
\multicolumn{1}{|c|}{} & 0.00 & 0.00 & 0.00 & 0.00 & 0.00 & 0.00 & 0.00 & 0.00 & 0.00 & 0.00 & 0.00 & 0.00 \\
\hline

\multicolumn{1}{|c|}{\multirow{2}{*}{2021}} & N/A & N/A & N/A & N/A & N/A & N/A & N/A & N/A & N/A & N/A & N/A & N \\
\cline{2-13}
\multicolumn{1}{|c|}{} & N/A & N/A & N/A & N/A & N/A & N/A & N/A & N/A & N/A & N/A & N/A & 0.00 \\
\hline

\multicolumn{13}{|c|}{Account Closure Date: N/A} \\
\hline
\end{tabular}
}
\captionof{table}{Revolving Credit Account Details}
\label{tab:credit_account_revolving2}
\end{minipage}
\end{table}

\begin{table}[p]
\centering
\renewcommand\arraystretch{1.2}
\begin{minipage}{\textwidth}
\centering
\begin{tabular}{|c|c|c|c|c|}
\hline
\multicolumn{5}{|c|}{\textbf{Residence Information}} \\
\hline
ID & Address & Phone & Status & Update Date \\
\hline
1 & Room 502, Unit 2, Building 3, No. XXX  Zhengzhou City & 0371—67845621 & Rented & 2023.08.15 \\
\hline
\end{tabular}
\captionof{table}{Residence Information Table Details}
\label{tab:residence_info}
\end{minipage}
\end{table}
\clearpage

\begin{table}[p]
\centering
\renewcommand\arraystretch{1.2}
\begin{tabular}{|c|c|c|c|c|}
\hline
\multicolumn{5}{|c|}{\textbf{Credit Agreement 1}} \\
\hline
Management Institution & Credit Agreement ID & Effective Date & Maturity Date & Credit Limit Purpose \\
\hline
Commercial Bank ZZ & E705 & 2021.09.15 & 2024.08.31 & Non-revolving Loan Limit \\
\hline
Credit Limit & Credit Quota & Credit Quota ID & Used Limit & Currency \\
\hline
71747 & 71747 & M806 & 19831 & CNY \\
\hline
\end{tabular}
\end{table}
\begin{table}[p]
\centering
\renewcommand\arraystretch{1.2}
\begin{tabular}{|c|c|c|c|c|}
\hline
\multicolumn{5}{|c|}{\textbf{Credit Agreement 2}} \\
\hline
Management Institution & Credit Agreement ID & Effective Date & Maturity Date & Credit Limit Purpose \\
\hline
Commercial Bank HN & E357 & 2023.08.20 & 2025.03.15 & Revolving Loan Limit \\
\hline
Credit Limit & Credit Quota & Credit Quota ID & Used Limit & Currency \\
\hline
15000 & 15000 & M953 & 0 & CNY \\
\hline
\end{tabular}
\end{table}
\begin{table}[p]
\centering
\renewcommand\arraystretch{1.2}
\begin{tabular}{|c|c|c|c|c|}
\hline
\multicolumn{5}{|c|}{\textbf{Credit Agreement 3}} \\
\hline
Management Institution & Credit Agreement ID & Effective Date & Maturity Date & Credit Limit Purpose \\
\hline
Commercial Bank JR & E788 & 2024.11.28 & 2025.10.31 & Revolving Loan Limit \\
\hline
Credit Limit & Credit Quota & Credit Quota ID & Used Limit & Currency \\
\hline
96148 & 96148 & M626 & 7993 & CNY \\
\hline
\end{tabular}
\end{table}
\begin{table}[p]
\centering
\renewcommand\arraystretch{1.2}
\begin{tabular}{|c|c|c|c|c|}
\hline
\multicolumn{5}{|c|}{\textbf{Credit Agreement 4}} \\
\hline
Management Institution & Credit Agreement ID & Effective Date & Maturity Date & Credit Limit Purpose \\
\hline
ICBC Zhengzhou Branch & E680 & 2021.09.15 & 2024.12.31 & Credit Card Shared Limit \\
\hline
Credit Limit & Credit Quota & Credit Quota ID & Used Limit & Currency \\
\hline
15000 & 15000 & M461 & 12800 & CNY \\
\hline
\end{tabular}
\end{table}
\begin{table}[p]
\centering
\renewcommand\arraystretch{1.2}
\begin{tabular}{|c|c|c|c|c|}
\hline
\multicolumn{5}{|c|}{\textbf{Credit Agreement 5}} \\
\hline
Management Institution & Agreement ID & Effective Date & Maturity Date & Credit Limit Purpose \\
\hline
Zhengzhou Bank & E326 & 2021.09.15 & 2024.12.31 & Credit Card Independent Limit \\
\hline
Credit Limit & Credit Quota & Credit Quota ID & Used Limit & Currency \\
\hline
15000 & 15000 & M952 & 12800 & CNY \\
\hline
\end{tabular}
\end{table}
\begin{table}[p]
\centering
\renewcommand\arraystretch{1.2}
\begin{tabular}{|c|c|c|c|c|}
\hline
\multicolumn{5}{|c|}{\textbf{Credit Agreement 6}} \\
\hline
Management Institution & Agreement ID & Effective Date & Maturity Date & Credit Limit Purpose \\
\hline
CCB Zhengzhou Jinshui Branch & E287 & 2023.03.20 & 2024.12.31 & Credit Card Shared Limit \\
\hline
Credit Limit & Credit Quota & Credit Quota ID & Used Limit & Currency \\
\hline
8000 & 8000 & M064 & 7200 & CNY \\
\hline
\end{tabular}
\end{table}
\begin{table}[p]
\centering
\renewcommand\arraystretch{1.2}
\begin{minipage}{\textwidth}
\centering
\begin{tabular}{|c|c|c|c|c|}
\hline
\multicolumn{5}{|c|}{\textbf{Credit Agreement 7}} \\
\hline
Management Institution & Agreement ID & Effective Date & Maturity Date & Credit Limit Purpose \\
\hline
CCB Jinshui Branch & E468 & 2023.03.20 & 2024.12.31 & Credit Card Independent Limit \\
\hline
Credit Limit & Credit Quota & Credit Quota ID & Used Limit & Currency \\
\hline
8000 & 8000 & M526 & 7200 & CNY \\
\hline
\end{tabular}
\captionof{table}{Credit Agreement Tables Details.}
\label{tab:credit_agreement_7}
\end{minipage}
\end{table}
\end{document}